\RequirePackage{lineno} 

\documentclass[twocolumn,preprintnumbers,amsmath,amssymb,nofootinbib]{revtex4}

\usepackage{graphicx}
\usepackage{dcolumn}
\usepackage{bm}
 
\usepackage{epstopdf}

\def\bea{\begin{eqnarray}}
\def\eea{\end{eqnarray}}

\def\pp{\mbox{$p$-$p$}}

\def\pa{\mbox{$p$-A}}

\def\auau{\mbox{Au-Au}}

\def\pbpb{\mbox{Pb-Pb}}
\def\ppb{\mbox{$p$-Pb}}

\def\pn{\mbox{$p$-N}}
\def\aa{\mbox{A-A}}
\def\nn{\mbox{N-N}}

\def\ee{\mbox{$e^+$-$e^-$}}

\def\pt{$p_t$}

\def\yt{$y_t$}

\def\nch{$n_{ch}$}
\def\mmpt{$\bar p_t$}

\begin{document} 

\setlength{\pdfpagewidth}{8.5in}
\setlength{\pdfpageheight}{11in}

\setpagewiselinenumbers
\modulolinenumbers[5]

\preprint{version 3.6}

\title{Space-time geometry of small and large collision systems at LHC energies
}

\author{Thomas A.\ Trainor}\affiliation{University of Washington, Seattle, WA 98195}


\date{\today}

\begin{abstract}
Identified-hadron (PID) spectra from 2.76 TeV Pb-Pb and $p$-$p$ collisions are analyzed via a two-component (soft + hard) model (TCM) of hadron production in high-energy nuclear collisions. The Pb-Pb TCM is adopted with minor changes from a recent analysis of PID hadron spectra from 5 TeV $p$-Pb collisions. The object of study is evidence for jet suppression in small and large collision systems as indicating quark-gluon plasma (QGP) formation there. Conventional methods have included Pb-Pb centrality determination via classical Glauber model and evidence for high-$p_t$ suppression sought via spectrum ratio $R_\text{AA}$. For the latter it is implicitly assumed that the jet contribution to spectra is factorizable, including as a factor number of binary N-N collisions $N_{bin}$. In previous $p$-Pb studies validity of the Glauber model for nuclear collisions is questioned. In the present study alternative geometry determination via ensemble-mean $\bar p_t$ data reveals that the number of participant nucleons in central Pb-Pb collisions is about 1/3 of the Glauber estimate. Based on certain features of Pb-Pb spectra the validity of the factorization assumption is also questioned. The entire jet contribution is therefore treated without factorization in ratio to a \mbox{$p$-$p$} spectrum model as reference. The new results indicate that exclusivity (a nucleon may only interact with one nucleon ``at a time'') and time dilation (experienced by participant partons) play an essential role in jet production not incorporated in Glauber model or hard component factorization. The combination determines an effective number of N-N collisions per participant nucleon given specific Pb-Pb centrality: multiple collisions if associated with low-$x$ (slow) partons, a single collision if associated with high-$x$ (fast) partons experiencing strong time dilation. The effect on parton fragment (jet) distributions on $p_t$ is similar to projectile-proton fragment distributions on pseudorapidity from fixed-target $p$-A experiments where low-$\eta$ densities scale with A while high-$\eta$ densities are consistent with $p$-$p$ collisions. $p$-Pb and Pb-Pb spectra similarly analyzed reflect the same physics given different geometries. Jet suppression related to QGP formation is not evident.
\end{abstract}

\maketitle

 \section{Introduction}
 
This article reports analysis of identified-hadron (PID) \pt\ spectra from six centrality classes of 2.76 TeV \pbpb\ collisions~\cite{alicepbpbpidspec} via a  two-component (soft + hard) model (TCM) of hadron production in high-energy nuclear collisions~\cite{ppprd,ppquad,hardspec,alicetomspec,anomalous,tommpt}. Data from 2.76 TeV \pp\ collisions~\cite{alicespec2}
are also included.  This analysis was intended to explore the centrality systematics of jet modification (jet quenching) in \pbpb\ collisions and test the hypothesis that quark-gluon plasma (QGP), a dense flowing QCD medium, is formed in \aa\ collisions. Such studies have been conventionally conducted with spectrum ratio $R_\text{AA}$ as defined in Eq.~(\ref{raa}) or central-peripheral measure $R_\text{CP}$. Reduction of such ratios below 1 at higher \pt\ has been interpreted to indicate jet modification or ``quenching.''
  
 An alternative approach addresses accurate representation of {\em any} information carried by spectrum data. An example is given by Ref.~\cite{ppprd} describing a process in which a two-component spectrum model is inferred from the charge multiplicity (\nch) dependence of \pt\ spectra from 200 GeV \pp\ collisions. The physical origins of the two spectrum components have been inferred by comparisons with QCD theory~\cite{fragevo} and correlation data~\cite{porter2,porter3}. 
In contrast to full-spectrum ratios such as $R_\text{AA}$ the TCM accurately separates soft (nonjet) and hard (jet) contributions to spectra. Such separation permits examination of entire jet fragment distributions for possible modification, not just the higher-\pt\ parts of spectra.
 The TCM has been applied successfully to 200 GeV \auau\ collisions~\cite{hardspec}, a large volume of \pp\ data over several collision energies~\cite{alicetomspec}, \ppb\ PID spectrum data~\cite{ppbpid} and two-particle angular correlations~\cite{anomalous,ppquad}.  As a {\em composite}  production model the TCM is apparently required by spectrum~\cite{alicetomspec} and correlation~\cite{ppquad} data for all A-B collision systems, and the two components overlap strongly on \pt.  The present study is an extension of that program to determine the properties of \pbpb\ PID \pt\ spectra.

An essential part of spectrum interpretation is determination of collision geometry, taken here to mean estimation of number of nucleon participants $N_{part}$ and nucleon-nucleon binary collisions $N_{bin}$ for various event classes. A conventional method relies on classical Glauber Monte Carlos, but that approach encounters major difficulties with \ppb\ spectra as reported in Refs.~\cite{ppbpid,tomglauber}. An alternative method derives geometry information from ensemble \mmpt\ data. There are large differences between the two methods for \pbpb\ geometry as described below.

With collision geometry established via \mmpt\ data the \pbpb\ PID spectra are processed according to the standard TCM. Two notable results emerge: (a) when published spectra are rescaled by a soft density $\bar \rho_{si}$ (for hadron species $i$), determined via the TCM and \mmpt\ data, \pbpb\ spectra above a certain \pt\ value {\em coincide  with \pp\ spectra} for all event classes and (b) isolated hard components {\em appear} suppressed (relative to a model) above a certain \pt\ value but show no evidence for such suppression {\em below} that point. Those results call into question the conventional concept of jet suppression.

In response to those observations the usual factorization assumed for the hard component within  the TCM is set aside. Entire hard components are retained intact and \pbpb\ hard components are compared directly with \pp\ hard components. What emerges strongly suggests that a combination of exclusivity~\cite{tomexclude} and time dilation are responsible for what has been interpreted as jet suppression in conventional analyses. There is strong similarity with certain results from \pa\ experiments in the seventies interpreted to reveal {\em nuclear transparency}~\cite{buszanew}.

Previous detailed studies of 5 TeV \ppb\ PID spectra~\cite{ppbpid,pidpart1,pidpart2} reported small variations of hard-component peak modes (positions of maxima) and peak widths, but results were inconclusive as to jet suppression or not. Application of  methods developed for \pbpb\ spectra in the present study reveal {\em quantitatively} different but similar behavior in \ppb\ data as in \pbpb\ data. Exclusivity and time dilation play a dominant role. The main difference between the two systems is collision geometry, not the presence or absence of a dense medium. These results have major implications for data interpretations based on formation of a dense QCD medium with consequent jet suppression and should motivate reexamination of the consequences of exclusivity and time dilation for jet production within larger target nuclei.

This article is arranged as follows:
Section~\ref{alicedata} presents 2.76 TeV  \pbpb\ PID spectrum data from Ref.~\cite{alicepbpbpidspec} in a conventional presentation format.
Section~\ref{spectrumtcm} describes an A-B collision PID spectrum TCM.
Section~\ref{geometry} reviews different methods of determining collision geometry.
Section~\ref{pidspec} reports a 2.76 TeV  \pbpb\ PID spectrum TCM resulting from the present study.
Section~\ref{extend} extends the TCM to a form not assuming hard-component factorization.
Section~\ref{ratios} compares spectrum ratios in various formats.
Section~\ref{sys}  reviews systematic uncertainties.
Sections~\ref{disc} and~\ref{summ} present discussion and summary.

\section{2.76 $\bf TeV$ $\bf Pb$-$\bf Pb$ PID Spectrum data} \label{alicedata}

Conventional approaches to analysis of \pt\ spectra from high-energy nuclear collisions have been motivated by an adopted narrative in which partial equilibration of flowing QCD matter at the parton level occurs within the collision space-time volume, and most detected hadrons emerge from a low-viscosity flowing medium or ``perfect liquid''~\cite{perfect} via a subsequent bulk hadronization process. Several statements in Ref.~\cite{alicepbpbpidspec} illustrate this narrative:

``In...heavy ion collisions, a strongly-interacting deconfined medium of quarks and gluons is created.''
``Transverse  momentum...spectra probe many different properties of this medium.''
Flows are assumed to dominate the low-\pt\ part of a spectrum (e.g.\ below 2-3 GeV/c) while jet fragments (assumed to be a small fraction of all hadrons) should dominate at higher \pt\ (e.g.\ above 6-10 GeV/c). 
The \pt\ spectra of hadrons should then reveal a velocity profile of the flowing dense medium and possible modification of jet formation due to passage of energetic scattered partons through the conjectured medium. 
In the lower \pt\ interval spectrum features are expected to reflect ``bulk production.'' So-called ``hardening of the spectra [with increasing centrality] is mass dependent and is characteristic of hydrodynamic flow....'' 
A blast-wave model  fitted to spectra over a limited lower-\pt\ interval includes parameters interpreted to represent a ``kinetic freezeout'' temperature and a radial flow velocity~\cite{blastwave}. 

In the higher-\pt\ interval jet formation and jet modification are anticipated as the relevant issues.  
``...jet  quenching leads to a suppression of high-\pt\ particle production...''
``[T]he spectra follow a power-law shape as expected from perturbative QCD (pQCD) calculations.''
Spectrum ratio $R_\text{AA}$ is interpreted to reveal the effects of jet quenching~\cite{jetquenching} as reductions from unity (jet suppression) at higher \pt~\cite{raastar}. 
``The microscopic mechanism of jet quenching is not completely understood, and one of the main goals of the experimental programs [at RHIC and LHC] is [to further understand jet quenching].''

Spectrum trends in the intermediate-\pt\ region (e.g.\ the ``baryon/meson puzzle'' \cite{rudy,duke,tamu}) appear to be still not clearly understood.
``Recent measurements...in \ppb\ collisions have revealed phenomena typically associated with fluid-like behavior in [\aa] collisions. ... The centrality evolution ... for \pbpb\ collisions can therefore also be seen as a possible experimental interconnection between the smallest and the largest QCD bulk systems.''
The paper concludes that ``...the centrality dependence presented in  this paper is important for constraining both the low-\pt\ hydrodynamics and the high-\pt\ jet quenching in the [theoretical model] calculations.''

\subsection{Pb-Pb PID  spectrum data} \label{piddataa}

Identified-hadron spectrum data for the present analysis reported in Ref.~\cite{alicepbpbpidspec} were produced by the ALICE collaboration at the LHC. A major goal of that study was determination of nuclear modification factors (NMFs) out to $p_t \approx 20$ GeV/c. The azimuth acceptance is $2\pi$ and pseudorapidity acceptance is $|\eta|< 0.8$ or $\Delta \eta = 1.6$.  \pbpb\ Collisions ($40\times 10^6$ events) are divided into six centrality classes: 0-5\%, 5-10\%, 10-20\%, 20-40\%, 40-60\% and 60-80\%. Corresponding estimated centrality parameters from Ref.~\cite{pbpbcent} are presented in Table \ref{ppbparams1}.
Hadron species include charged pions $\pi^\pm$, charged kaons $K^\pm$ and protons $p,~\bar p$. The \pp\ spectra ($11\times 10^6$ events) are reported in Ref.~\cite{alicespec2}. 

\begin{figure}[h]
	\includegraphics[width=1.65in]{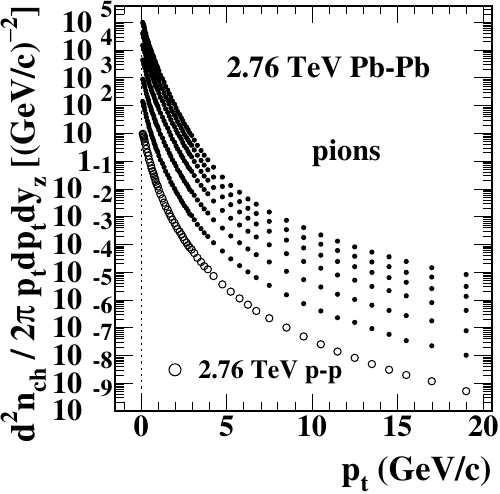}
	\includegraphics[width=1.65in]{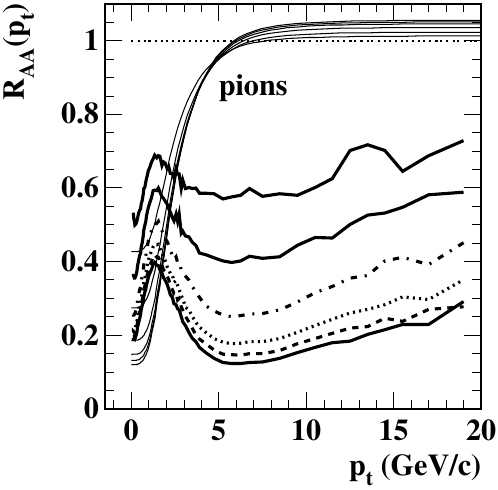}
\put(-142,85) {\bf (a)}
\put(-21,95) {\bf (d)}\\
	\includegraphics[width=1.65in]{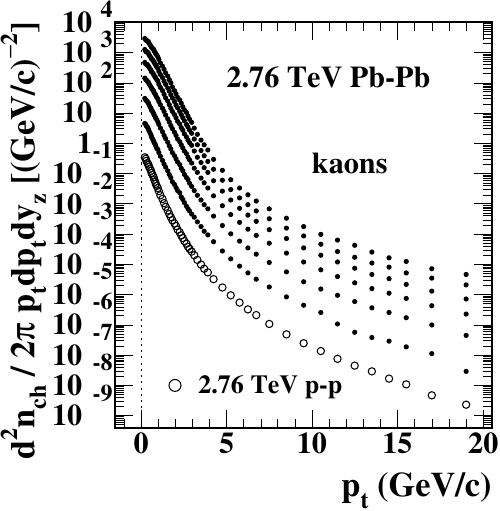}
\includegraphics[width=1.65in]{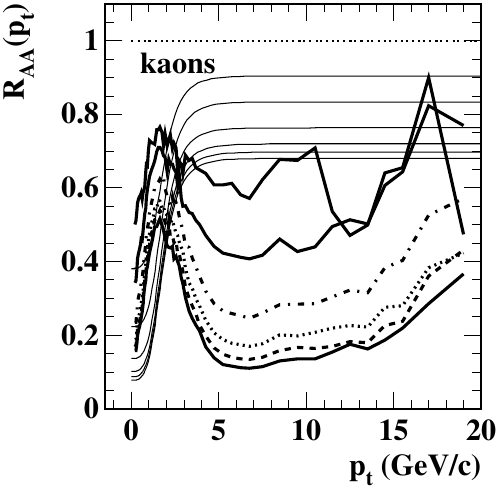}
\put(-142,85) {\bf (b)}
\put(-21,32) {\bf (e)}\\
	\includegraphics[width=1.65in]{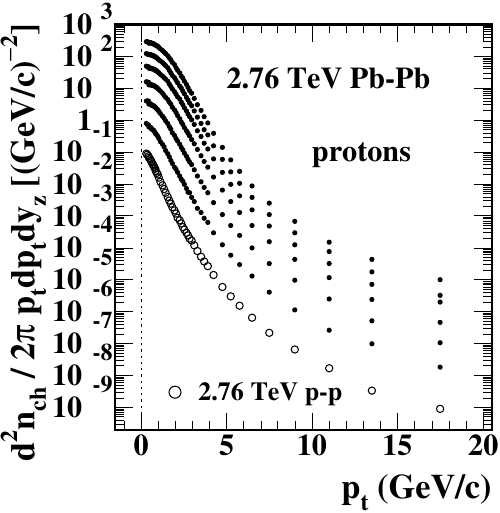}
\includegraphics[width=1.65in]{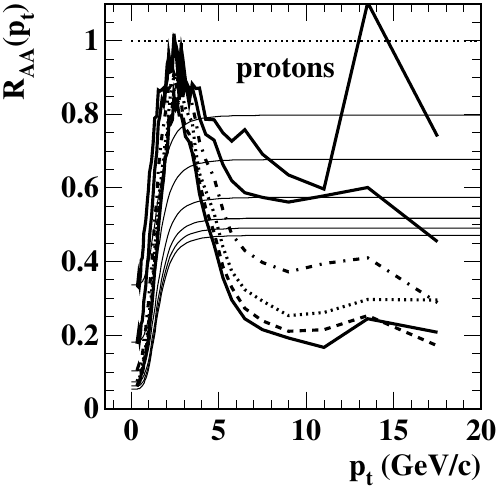}
\put(-142,85) {\bf (c)}
\put(-20,97) {\bf (f)}\\
	\caption{\label{piddata}
Left:
\pt\ spectra for three species of identified hadrons from six centrality classes of 2.76 TeV \pbpb\ collisions (solid) and from \pp\ collisions (open)~\cite{alicepbpbpidspec}.
Right: Spectrum ratios (nuclear modification factors) $R_\text{AA}(p_t)$ derived from data in the left panels (bold curves of various line styles). The thin solid reference curves are described below Eq.~(\ref{raann}).
}  
\end{figure}

Figure~\ref{piddata} (a-c) shows PID spectra  $\bar \rho_\text{0PbPbi}(p_t)$ and $\bar \rho_{0ppi}(p_t)$ from Ref.~\cite{alicepbpbpidspec} (points) in a conventional  semilog plotting format vs linear hadron \pt. Index $i$ represents a specific hadron species. The spectra are scaled by powers of 2 according to $2^{m-1}$ where $m \in [1,6]$ and $m = 6$ is most central (not to be confused with centrality index $n \in [1,7]$ for \ppb\ collisions in Ref.~\cite{ppbpid}). The selected plot format greatly limits visual access to {\em differential} spectrum features as they vary with hadron species and \pbpb\ centrality, especially at lower \pt\ where {\em most jet fragments appear}~\cite{ppprd,fragevo}. Compare with corresponding figures presented in a TCM format in Sec.~\ref{pidspec}.

Figure~\ref{piddata} (d-f) show spectrum ratios $R_\text{AA}$  (nuclear modification factors or NMFs) of \pbpb\ spectra vs \pp\ spectra rescaled by the number of \nn\ binary collisions $N_{bin}$ estimated by a Glauber Monte Carlo as in Table~\ref{ppbparams1}~\cite{pbpbcent},
\bea \label{raa}
R_\text{AAi}(p_t,n_{ch}) &=& \frac{1}{N_{bin}}\frac{\bar \rho_\text{0PbPbi}(p_t,n_{ch})}{\bar \rho_{0ppi}(p_t)}.
\eea

For the ideal or reference case that \aa\ collisions can be described as {\em linear superpositions} of \nn\ $\approx$ \pp\ collisions the expression in Eq.~(\ref{raa}) can be idealized to obtain
\bea  \label{raann}
R_{\text{AA}i}(y_t) &\rightarrow& \frac{1}{N_{bin}} \frac{(N_{part}/2)\bar \rho_{sNNi}\hat S_{0i} + N_{bin} \bar \rho_{hNNi}\hat H_{0i}}{\bar \rho_{sppi}\hat S_{0i} +  \bar \rho_{hppi}\hat H_{0i}}
\nonumber \\
&& \hskip -.3in =~\frac{1}{\nu} \frac{1 + \tilde z_{i} x_{NN} \nu \, \hat T_{0i}(y_t)}{1 + \tilde z_{i}x_{pp}\hat T_{0i}(y_t)} \times \frac{z_{siAA}(\nu)}{z_{sipp}},
\eea
(see PID spectrum definition in Sec.~\ref{spectrumtcm}) where the model functions are unmodified (fixed),  $\hat T_{0i}(y_t) \equiv \hat H_{0i}(y_t) / \hat S_{0i}(y_t)$ and $x_{NN} \equiv \bar \rho_{hNN} / \bar \rho_{sNN}\approx x_{pp} \approx \alpha \bar \rho_{spp}$, with $\tilde z_{i}x_{NN}(n_s)\hat T_0(y_t) \ll 1$ for low \yt\ and $\gg 1$ for high \yt~\cite{alicetomspec}. The ratio of $z_{si}$ factors is determined by  Eq.~(\ref{rhosi}). 

Equation~(\ref{raann}) generates the thin solid TCM reference curves in Fig.~\ref{piddata} (right). Asymptotic values are $1/\nu = N_{part}/2N_{bin}$ at low \pt\ and 1 at high \pt, both multiplied by AA/\pp\ soft-fraction ratio $z_{siAA}(\nu)/z_{sipp}$. The last implies that $R_\text{AA}(y_t)$ as conventionally interpreted would signal substantial ``jet suppression'' for heavier hadron species (i.e.\ baryons) {\em in the absence of any dense medium}. In panel (f) the TCM reference accurately describes particularly low values of proton $R_\text{AA}(y_t)$ data at low \pt.

The independent variable in these plots is transverse momentum \pt. The majority of the plot space, on the right, is then occupied by slowly-varying data representing less than 7\% of total jet fragments while rapid changes involving {\em most} jet fragments are compressed into a small interval on the left (the narrow peaks on $p_t$).
Transverse rapidity $y_{t\pi} \equiv \ln[(p_t + m_{t\pi})/m_\pi]$ is an alternative  variable assuming pion mass for all hadron species. 
As such, $y_{t\pi}$ is simply a logarithmic representation of transverse momentum \pt\ with a well-defined zero. The logarithmic momentum scale gives much-improved visual access to the lower-\pt\ large jet contributions~\cite{fragevo,alicetomspec,ppquad}. 

\subsection{$\bf Pb$-Pb Glauber-model geometry parameters} \label{pbpbgeomx}

In order to proceed with $R_\text{AA}$ estimation and other aspects of spectrum analysis collision geometry parameters $N_{part}$ (number of nucleon participants N) and $N_{bin}$ (number of {\em binary} \nn\ collisions) must be estimated. This is conventionally accomplished with a classical Glauber Monte Carlo model of the \pbpb\ collision process~\cite{pbpbcent}.

Table~\ref{ppbparams1} shows geometry parameters for 2.76 TeV \pbpb\ collisions from Table~1 of Ref.~\cite{pbpbcent} inferred from a Glauber Monte Carlo corresponding to the spectra in Fig.~\ref{piddata}. Certain power-law relations among Glauber parameters and charge density $\bar \rho_0 = n_{ch} / \Delta \eta$ are notable.

\begin{table}[h]
	\caption{Glauber parameters for 2.76 TeV \pbpb\ collisions from Table 1 of Ref.~\cite{pbpbcent} (primed entries). Also included are values from the analysis in Sec.~\ref{altgeom} (unprimed entries). $\nu$ values are $x\nu$ values (solid squares) from Fig.~\ref{inverse} divided by constant value $x = 0.163$.  $\bar \rho_s = \bar \rho_0/(1+x\nu)$ and $\bar \rho_0$ values are from Table 1 of Ref.~\cite{pbpbcent}. $N_{part}(/2)$ values are from open squares in Fig.~\ref{story} (b). Note that for central \pbpb\ $N_{part} = 140$ is only 1/3 of the total number of nucleons in Pb+Pb.
	}
	\label{ppbparams1}
	\begin{center}
		\begin{tabular}{|c|c|c|c|c|c|c|c|} \hline
			centrality & $b_\text{max}'$ (fm) & $N_{part}'$ & $N_{bin}'$ & $\nu'$ & $\bar \rho_s$ & $\nu$  & $N_{part}$ \\ \hline
			0 - 5   &  3.5   & 383 & 1685 & 8.8 & 1092 & 2.9 & 140  \\ \hline
			~\,5 - 10   & 4.9  & 329& 1316  & 8.0 & 903 & 2.7 & 116  \\ \hline
			10 - 20   &7.0  & 260 & 921  &  7.1 & 686 & 2.5 & 89 \\ \hline
			20 - 40   & 9.9  & 157  & 438 &  5.6 & 402 &2.1 & 52 \\ \hline
			40 - 60     & 12.1  & 68.6 & 128  & 3.7  & 163 &1.6 & 21 \\ \hline
			60 - 80    & 14.0  &22.5 & 26.7  & 2.4 &46 & 1.2 & 6 \\ \hline
		\end{tabular}
	\end{center}
\end{table}

As context, a Pb nucleus is approximately eight tangent nucleons in diameter, and $\nu  \equiv 2 N_{bin} / N_{part}$ is the mean number of \nn\ binary collisions per \nn\ pair averaged over the \pbpb\ overlap region. For central \pbpb\ collisions $\nu \approx 9$ implies that on average a ``projectile'' nucleon must collide each time {\em simultaneously} with 2-3 ``target'' nucleons. A study of \ppb\ geometry in Ref.~\cite{tomglauber} and follow-up study of Glauber-model inconsistencies in Ref.~\cite{tomexclude} suggest that at least in the asymmetric \ppb\ context multiple {\em simultaneous} \pn\ collisions are prohibited. To proceed with the \pbpb\ geometry it is useful to identify some relationships among Glauber parameters.

\begin{figure}[h]
	\includegraphics[width=1.65in]{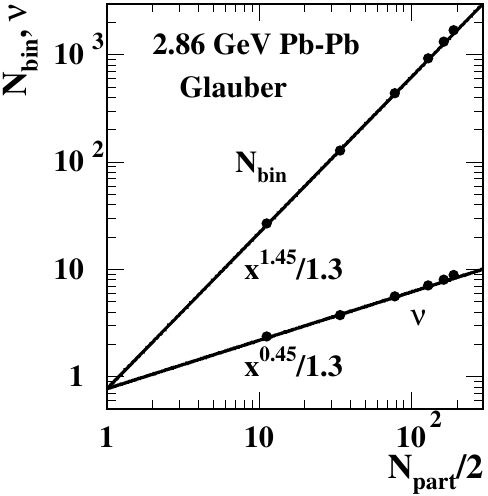}
	\includegraphics[width=1.65in]{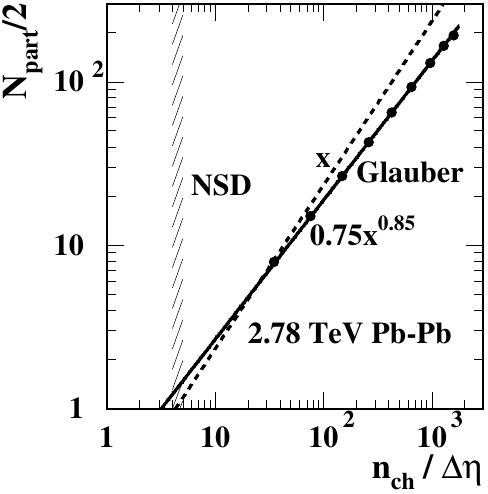}
	\caption{\label{glauber}
		Left: Glauber Monte Carlo values of $N_{part}/2$ and $N_{bin}$ from Table~\ref{ppbparams1} illustrating power-law relationships.
		Right: Empirical power-law relation between $N_{part}/2$ and charge density $\bar \rho_0$. See Fig.~\ref{story} (b). The dashed line is a reference.
	}  
\end{figure}

Figure~\ref{glauber} (left) shows power-law relations (solid points) among parameters $N_{part}$, $N_{bin}$ and $\nu$ inferred from classical-Glauber parameters in Table~\ref{ppbparams1} (primed values). Specifically, $N_{bin} \propto N_{part}^{1.45}$ and $\nu  \propto N_{part}^{0.45}$. Those relations correspond to the basic geometry of two interpenetrating spheres. 
Figure~\ref{glauber} (right) shows an empirical relation (solid curve) between $N_{part}/2$ and charge density $\bar \rho_0$  based on Glauber-related data in Ref.~\cite{alicepbpbyields}. The dashed line is $N_{part}/2 = \bar \rho_0/\bar \rho_{sNN}$ with $\bar \rho_{sNN} = 4.3$.

As demonstrated in Fig.~\ref{900a} (left) the power-law trends in this figure predict a nonphysical $2\bar \rho_0/N_{part}$ trend (bold dotted curve in that figure) that bypasses the \pp\ value $\bar \rho_0 \approx 4.55$ for $\nu =1$. An alternative TCM-related determination based on \mmpt\ data is presented in Sec.~\ref{altgeom}.

\section{A-B collisions PID Spectrum $\bf TCM$} \label{spectrumtcm}

High-energy nuclear data exhibit a  basic property:  data distributions are typically {\em composite} and require a two-component (soft + hard) model of hadron production as demonstrated initially for \pt\ spectra from \pp\ collisions~\cite{ppprd,ppquad}. The two distinct components overlap strongly on \pt, in contrast to conventional assumptions about spectrum structure described above. PID spectra from 2.76 TeV \pp\ and \pbpb\ collisions are analyzed in this study in the context of previous analysis of \pp~\cite{ppprd,ppquad},  and \ppb~\cite{tomglauber,ppbpid} data which provide reference trends. The TCM for \pp\ and \ppb\  collisions is the product of phenomenological analysis of data from multiple collision systems and data formats~\cite{ppprd,hardspec,ppquad,alicetomspec,tommpt}. Physical interpretations of TCM soft and hard components have been derived {\em a posteriori} by comparing inferred TCM characteristics with other relevant measurements~\cite{hardspec,fragevo}, in particular measured minimum-bias (MB) dijet properties~\cite{eeprd,jetspec2,mbdijets}. The TCM does not result from fits to individual spectra which would require many parameter values. The few A-B TCM parameters are required to have simple $\log(\sqrt{s})$ trends on collision energy and simple extrapolations from measured \pp\ trends.

To establish a TCM for A-B PID \pt\ spectra it is assumed that (a) \nn\ parameters $\alpha$, $\bar \rho_{sNN}$ and $\bar \rho_{hNN}$ have been inferred from non-PID data and (b) geometry parameters $N_{part}$, $N_{bin}$, $\nu$ and $x$ for a specific A-B collision system  are a common property (relating to centrality) of any hadron species within that system. In this section geometry parameters determined by a Glauber Monte Carlo are briefly summarized as an introduction. Determination of geometry parameters by two different methods is examined in more detail in Sec.~\ref{geometry} below.

\subsection{A-B spectrum TCM for unidentified hadrons} \label{unidspec}

The \pt\ or \yt\ spectrum TCM is by definition the sum of distinct soft and hard components with details inferred from data (e.g.\ Ref.~\cite{ppprd}). For \pp\ collisions
\bea  \label{rhotcm}
\bar \rho_{0}(y_t,n_{ch}) &=& S_{pp}(y_t,n_{ch}) + H_{pp}(y_t,n_{ch})
\\ \nonumber
&\approx& \bar \rho_{s}(n_{ch}) \hat S_{0}(y_t) + \bar \rho_{h}(n_{ch}) \hat H_{0}(y_t),
\eea
where hadron charge \nch\ is here an event-class index, and factorization of  dependences on \yt\ and \nch\ is a main feature of the spectrum TCM. The motivation for transverse rapidity $y_{ti} \equiv \ln[(p_t + m_{ti})/m_i]$ (applied to hadron species $i$) is described in Sec.~\ref{tcmmodel}. The \yt\ integral of Eq.~(\ref{rhotcm}) is $\bar \rho_0 \equiv n_{ch} / \Delta \eta = \bar \rho_s + \bar \rho_h$, a sum of soft and hard charge densities. $\hat S_{0}(y_t)$ and $\hat H_{0}(y_t)$ are unit-normal model functions approximately independent of \nch, and the centrally-important relation $\bar \rho_{h} \approx \alpha \bar \rho_{s}^2$ with $\alpha \approx O(0.01)$ is inferred from \pp\ spectrum data~\cite{ppprd,ppquad,alicetomspec}. The \pp\ hard/soft ratio is denoted by $x = \bar \rho_h / \bar \rho_s \approx \alpha \bar \rho_s$.

For composite A-B collisions the TCM is generalized to
\bea \label{ppspectcm}
\bar \rho_{0}(y_t,n_{ch}) 
&\approx& \frac{N_{part}}{2}  \bar \rho_{sAB} \hat S_{0} + N_{bin}  \bar \rho_{hAB} \hat H_{0}(n_{ch}),
\eea
including a further factorization of charge densities $\bar \rho_x = n_x / \Delta \eta$ into A-B geometry parameters $N_{part}$ (number of nucleon participants N) and $N_{bin}$ (\nn\ binary collisions) and mean charge densities $\bar \rho_{xAB}$  {\em per \nn\ pair} averaged over all \nn\ interactions within the A-B system. $\nu \equiv 2 N_{bin} / N_{part}$ is the number of binary collisions per participant pair.  For A-B collisions $\bar \rho_{s}(n_s) = [N_{part}(n_s)/2]\bar \rho_{sAB}(n_s)$ is a factorized soft-component density and $\bar \rho_h(n_s) = N_{bin}(n_s) \bar \rho_{hAB}(n_s)$ is treated within this description as a factorized hard-component density, with hard/soft ratio $x\nu = \bar \rho_h/ \bar \rho_s$.  Factorization of the hard component is addressed in more detail in Sec.~\ref{extend}.

\subsection{p-Pb geometry inferred from non-PID data} \label{geom}

Analysis of PID \ppb\ spectrum data reported in Ref.~\cite{ppbpid} provides a reference for \pp\ and \pbpb\ PID spectrum analysis in the present study. The \ppb\ geometry parameters from the earlier study are here reviewed.
Table~\ref{rppbdata} presents geometry parameters for 5 TeV \ppb\ collisions inferred from a Glauber-model analysis in Ref.~\cite{aliceglauber} (primed values) and TCM values from Ref.~\cite{tomglauber}  (unprimed values). Large differences between \ppb\ Glauber and TCM values are explained in Ref.~\cite{tomglauber}. Primed quantities based on a Glauber Monte Carlo study result from assumptions inconsistent with \ppb\ \mmpt\ data~\cite{tommpt}. Charge densities $\bar \rho_0$ are derived directly from data and correctly characterize the seven event classes.

\begin{table}[h]
	\caption{Glauber (primed~\cite{aliceglauber}) and  TCM (unprimed~\cite{tomglauber}) fractional cross sections and  geometry parameters, midrapidity charge density $\bar \rho_0$, \nn\ soft component $\bar \rho_{sNN}$ and TCM hard/soft ratio $x(n_s)$ used for 5 TeV \ppb\ PID spectrum data.
	}
	\label{rppbdata}
	\begin{center}
		\begin{tabular}{|c|c|c|c|c|c|c|c|c|c|} \hline
		$n$&	$\sigma' / \sigma_0$ &   $\sigma / \sigma_0$    & $N_{bin}'$ & $N_{bin}$ & $\nu'$ & $\nu$ & $\bar \rho_0$ & $\bar \rho_{sNN}$ & $x(n_s)$ \\ \hline
	1	& 	0.025         & 0.15 & 14.7  & 3.20 & 1.87  & 1.52 & 44.6 & 16.6  & 0.188 \\ \hline
	2	& 	0.075  & 0.24 &  13.0   & 2.59  &  1.86 & 1.43 & 35.9 &15.9  & 0.180 \\ \hline
	3	& 	0.15  & 0.37 &  11.7 & 2.16 & 1.84 &  1.37 & 30.0  & 15.2  & 0.172 \\ \hline
	4	& 	0.30 & 0.58 &  9.4 & 1.70 & 1.80  & 1.26  & 23.0  & 14.1  & 0.159  \\ \hline
	5	& 	0.50   &0.80  & 6.42  & 1.31 & 1.73  & 1.13 & 15.8 &   12.1 & 0.137  \\ \hline
	6	& 	0.70  & 0.95 & 3.81  & 1.07 & 1.58  & 1.03  & 9.7  &  8.7 & 0.098 \\ \hline
	7	& 	0.90 & 0.99 &  1.94 & 1.00 & 1.32  & 1.00  &  4.4  & 4.2 &0.047  \\ \hline
		\end{tabular}
	\end{center}
\end{table}

TCM (unprimed) geometry parameters in Table~\ref{rppbdata}, derived from \ppb\ \pt\ spectrum and \mmpt\ data for unidentified hadrons~\cite{tommpt,tomglauber}, are assumed to be valid also for each identified-hadron species and are used unchanged to process \ppb\ PID spectrum data in Ref.~\cite{ppbpid}. 

\subsection{PID spectrum TCM} \label{pidtcm}

Given the A-B spectrum TCM for unidentified-hadron spectra in Eq.~(\ref{ppspectcm}) a corresponding TCM for identified hadrons can be generated by assuming that each hadron species $i$ comprises certain fractions of soft and hard TCM components denoted by $z_{si}$ and $z_{hi}$  (both $\leq 1$) and assumed {\em independent of \yt} (but not centrality). The PID spectrum TCM can then be expressed as
\bea \label{pidspectcm}
\bar \rho_{0i}(y_t)
&\approx& \frac{N_{part}}{2}  \bar \rho_{sNNi} \hat S_{0i}(y_t) + N_{bin}   \bar \rho_{hNNi} \hat H_{0i}(y_t) 
\nonumber \\
\frac{\bar \rho_{0i}(y_t)}{ \bar \rho_{si}} 
&=& \hat S_{0i}(y_t) +  \tilde z_i(n_s) x(n_s)\nu(n_s) \hat H_{0i}(y_t,n_s)
\\ \nonumber
&\equiv&   \text{X}_i(y_t),
\eea
where $\bar \rho_{xNNi} = z_{xi}(n_s) \bar \rho_{xNN}$ ($x$ = $s$  or $h$), $\tilde z_{i}(n_s) \equiv z_{hi}(n_s) / z_{si}(n_s)$ and  unit-integral model functions $\hat S_{0i}(y_t)$ and $\hat H_{0i}(y_t)$ depend on hadron species $i$.
For identified hadrons of species $i$ \ rescale factor $ \bar \rho_{si}$ in the second line can be re-expressed in terms of $ \bar \rho_{s} =(N_{part}/2)\bar \rho_{sNN}$, already inferred for unidentified hadrons, as
\bea \label{rhosi}
{\bar \rho_{si}} &=&  \frac{\bar \rho_{0i}}{1+\tilde z_i x\nu}
\\ \nonumber
&=&z_{si}(n_s)  \bar \rho_{s}, ~~\text{with}
\\ \nonumber
z_{si}(n_s)&\approx& \left[\frac{1 + x(n_s) \nu(n_s)}{1 + \tilde z_{i}(n_s) x(n_s) \nu(n_s)} \right]  {z_{0i}},
\eea 
and where $z_{0i} = \bar \rho_{0i} /  \bar \rho_{0}$ is the fractional abundance of species $i$ relative to total hadron density $\bar \rho_{0}$.

\begin{figure}[h]
	\includegraphics[width=3.3in]{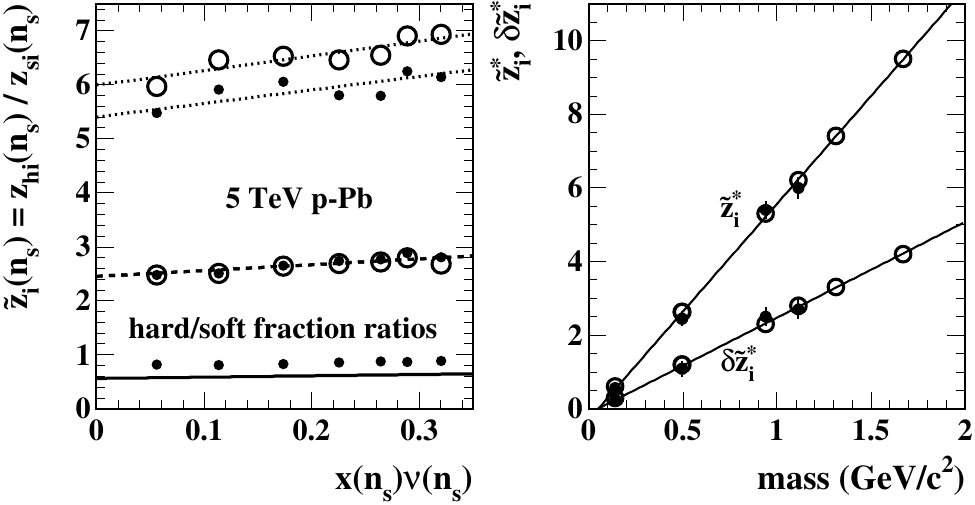}
	\caption{\label{tildezparams}
		Left: Ratios $\tilde z_{i}(n_s) = z_{hi}/z_{si}$ inferred from $z_{si}$ and $z_{hi}$ measurements reported in Ref.~\cite{pidpart1} for charged (solid dots) and neutral (open circles) hadrons. The lines are linear parametrizations $\tilde z_i = \tilde z_i^* + \delta \tilde z_i^* x(n_s) \nu(n_s)$ that describe the ratio data: solid, dashed and dotted for pions, kaons and baryons respectively.
		Right: Coefficients $\tilde z_i^*$ and $\delta \tilde z_i^*$ for linear descriptions of ratio data in the left panel plotted vs hadron mass. The lines represent proportionality to hadron mass. The solid dots are values inferred in Ref.~\cite{pidpart1}. Open circles provide linear extrapolations on hadron mass for $\Xi$ and $\Omega$.
	}  
\end{figure}

Figure~\ref{tildezparams} (left) shows ratios $\tilde z_{i}(n_s)$ (points) inferred from $z_{si}(n_s)$ and $z_{hi}(n_s)$ data reported in Ref.~\cite{pidpart1} for charged hadrons (solid dots) and neutral hadrons (open circles), from pions at the bottom to Lambdas at the top. Dashed and dotted lines corresponding to kaons and baryons are best fits by eye to the points with Eq.~(\ref{tildez}) 
\bea \label{tildez}
\tilde z_i(n_s) &=& \tilde z_i^* + \delta \tilde z_i^* x(n_s)\nu(n_s).
\eea
Parameters are presented as solid dots in the right panel.

Figure~\ref{tildezparams} (right) shows  $\tilde z_i^*$ and $\delta \tilde z_i^*$ (solid dots) for  Eq.~(\ref{tildez}) plotted vs hadron mass. Starred $\tilde z_i(n_s)$ parameters vary simply proportional to hadron mass and show no significant sensitivity to baryon identity or strangeness content. That empirical mass trend enables predictions of spectra for more-massive hadrons (e.g.\ Cascades and Omegas) from lower-mass pion, kaon and proton spectra~\cite{ppsss}. To simplify the present study $\tilde z_i$ values for midcentral collisions are adopted for all collision centralities.

\subsection{Spectrum TCM model functions} \label{tcmmodel}

Given rescaled spectrum data $\text{X}_i(y_t)$ as in Eq.~(\ref{pidspectcm}) and the trend $x(n_s) \sim n_s \sim n_{ch}$, spectrum soft component $\hat S_{0i}(y_t)$ is defined as the asymptotic limit of rescaled {\em data} spectra $X_i(y_t)$ as $n_{s} \rightarrow 0$. Hard components of data spectra are then defined as complementary to soft components. Thus, data soft and hard components are {\em inferred from data alone}, not imposed via assumed models.

The data soft component for specific hadron species $i$ is typically well described by a Boltzmann exponential on $m_{ti}  = \sqrt{p_t^2 + m_i^2}$ with power-law tail. The unit-integral soft-component model is 
\bea \label{s00}
\hat S_{0i}(m_{ti}) &=& \frac{A_i}{[1 + (m_{ti} - m_i) / n_i T_i]^{n_i}},
\eea
where $m_{ti}$ is transverse mass-energy for hadrons $i$ of mass $m_i$, $n_i$ is the power-law exponent, $T_i$ is the slope parameter and coefficient $A_i$ is determined by the unit-integral condition. Reference parameter values for unidentified hadrons from 5 TeV \pp\ collisions reported in Ref.~\cite{alicetomspec} are $(T,n) \approx (145~ \text{MeV},8.3)$. Model parameters $(T_i,n_i)$ for each species of identified hadrons as in Table~\ref{pidparams} are determined from \ppb\ spectrum data as described below.

Rescaled data spectra $\text{X}_i(y_t)$ per Eq.~(\ref{pidspectcm}) can be combined with model functions $\hat S_{0i}(m_t)$ and transformed to densities on \yt\ to extract {\em factorized} hard components as
\bea \label{pidhard}
Y_i(y_t) &\equiv& \frac{1}{\tilde z_i x(n_s) \nu(n_s) } \left[ \text{X}_i(y_t) -  \hat S_{0i}(y_t) \right].~~~
\eea
As defined by Eq.~(\ref{pidhard}) $\text{Y}_i(y_t)$ for \aa\ collisions is approximated by unit-normal  function $\hat H_{0i,AA}(y_t)$.  The hard-component model  is a Gaussian density on pion $y_{t\pi} \equiv \ln((p_t + m_{t\pi})/m_\pi)$ (as explained below) with exponential (on $y_t$) or power-law (on $p_t$) tail for larger \yt\
\bea \label{h00}
\hat H_{0}(y_t) &\approx & A \exp\left\{ - \frac{(y_t - \bar y_t)^2}{2 \sigma^2_{y_t}}\right\}~~~\text{near mode $\bar y_t$}
\\ \nonumber
&\propto &  \exp(- q y_t)~~~\text{for larger $y_t$ -- the tail},
\eea
where the transition from Gaussian to exponential on \yt\ is determined by slope matching~\cite{fragevo}.  As a density on \pt\ the $\hat H_0$ tail varies approximately as power law $1/p_t^{q + 2}$. Coefficient $A$ is determined by the unit-integral condition. Model parameters $(\bar y_t,\sigma_{y_t},q)$ for identified hadrons as in Table~\ref{pidparams} are also derived from \ppb\ spectrum data. A corresponding spectrum ratio $r_{\text{AA}i}(y_t)$ defined by
\bea \label{raax}
r_{\text{AA}i}(y_t) &\equiv& \frac{\text{Y}_i(y_t)}{\hat H_{0i,pp}'(y_t)} = \frac{\nu}{\nu'} \frac{x_\text{AA}}{x_{pp}'} \frac{ \hat H_{0i,\text{AA}}(y_t) }{\hat H_{0i,pp}'(y_t)}
\eea
where primed quantities are (possibly biased) {\em model} estimates and unprimed quantities are physical {\em data} values.  $r_{\text{AA}i}(y_t)$ is analogous to ratio $R_\text{AA}$ defined in Eq.~(\ref{raa}), but there are major differences between them.

In both expressions $\hat H_{0i,\text{AA}}(y_t)$ is a unit-normal representation of the \aa\ data hard component including nominal jet modification occurring in the \aa\ collision system. In ratio $r_{\text{AA}i}(y_t)$, $\hat H_{0i,pp}'(y_t)$ is a unit-normal TCM model function: a Gaussian + exponential tail as described above. However, in ratio $R_{\text{AA}i}(y_t)$ $\hat H_{0i,pp}(y_t)$ represents a physically measured \pp\ spectrum feature that may be subject to new physics or experimental bias. That difference may be assessed by replacing \aa\ spectra in $r_{\text{AA}i}(y_t)$ by a {\em measured} \pp\ spectrum to form $r_{ppi}(y_t)$.

All data spectra are plotted vs pion rapidity $y_{t\pi}$ with pion mass assumed.  
One motivation is comparison of spectrum hard components apparently arising from a common underlying jet spectrum on \pt~\cite{fragevo}, in which case $y_{t\pi}$ serves  as a logarithmic measure of hadron \pt\ with well-defined zero.  
Accuracy of TCM-data comparisons requires precise normalization of $\hat S_{0i}(m_{ti})$ and $\hat H_{0i}(y_{t})$.

Unit-normal model functions $\hat S_{0i}(m_t)$ and $\hat H_{0i}(y_t)$ for \pbpb\ and \pp\ data are retained from the \ppb\ analysis in Ref.~\cite{ppbpid}. As defined in Eq.~(\ref{s00}) $\hat S_{0i}(m_t)$ is a density on proper $m_{ti}$ for hadron species $i$ while $\hat H_{0i}(y_t)$ as defined in Eq.~(\ref{h00}) is  a density on pion $y_{t\pi}$ in all cases.

\subsection{5 TeV $\bf p$-$\bf Pb$ PID TCM spectrum parameters} \label{pidfracdata}
 
Equations~(\ref{ppspectcm}) - (\ref{h00}) with model parameters as in Tables~\ref{pidparams} and \ref{otherparams} below describe PID \pp\ and \ppb\ spectra within data uncertainties as reported in Refs~\cite{ppprd,ppquad,ppbpid} and serve as a {\em reference} for 2.76 TeV \pbpb\ PID spectrum data as in the present study.

Table~\ref{pidparams} shows TCM model parameters for hard component $\hat H_0(y_t)$ (first three) and soft component $\hat S_0(y_t)$ (last two). Hard-component model parameters vary slowly but significantly with hadron species. Centroids $\bar y_t$ shift to larger \yt\ with increasing hadron mass. Widths $\sigma_{y_t}$ are substantially larger for mesons than for baryons. Only $K_\text{S}^0$ and $\Lambda$ data extend to sufficiently high \pt\ to determine exponent $q$ which is substantially larger for baryons than for mesons. The combined centroid, width and exponent trends result in near coincidence among the several species for larger \yt. Evolution of TCM hard-component parameters with hadron species is consistent with measured PID FFs (refer to Fig.~7 of Ref.~\cite{ppbpid}) and with a common underlying parton (jet) spectrum for all TCM hard components.
  
 \begin{table}[h]
 	\caption{TCM model parameters for unidentified hadrons $h$ from Ref.~\cite{alicetomspec} and for identified hadrons from 5 TeV \ppb\ collisions from Ref.~\cite{ppbpid}: hard-component parameters $(\bar y_t,\sigma_{y_t},q)$ and soft-component parameters $(T,n)$. Numbers without uncertainties are adopted from a comparable hadron species with greater accuracy (e.g.\ $p$ vs $\Lambda$). 
 	}
 	\label{pidparams}
 	\begin{center}
 		\begin{tabular}{|c|c|c|c|c|c|} \hline
 			& $\bar y_t$ & $\sigma_{y_t}$ & $q$ & $T$ (MeV) &  $n$  \\ \hline
 			$ h $     &  $2.64\pm0.03$ & $0.57\pm0.03$ & $3.9\pm0.2$ & $145\pm3$ & $8.3\pm0.3$ \\ \hline
 			$ \pi^\pm $     &  $2.52\pm0.03$ & $0.56\pm0.03$ & $4.0\pm1$ & $145\pm3$ & $8.5\pm0.5$ \\ \hline
 			$K^\pm$    & $2.65$  & $0.58$ & $4.0$ & $200$ & $14$ \\ \hline
 			$K_\text{S}^0$          &  $2.65\pm0.03$ & $0.58\pm0.02$ & $4.0\pm0.2$ & $200\pm5$ & $14\pm2$ \\ \hline
 			$p$        & $2.92\pm0.02$  & $0.47$ & $4.8$ & $210\pm10$ & $14\pm4$ \\ \hline
 			$\Lambda$       & $2.96\pm0.02$  & $0.47\pm0.03$ & $4.8\pm0.5$  & $210$ & $14$ \\ \hline	
 		\end{tabular}
 	\end{center}
 \end{table}
  
 Soft-component model parameter $T \approx 145$ MeV for pions is the same as that for unidentified hadrons found to be universal over all A-B collision systems and collision energies~\cite{alicetomspec}. The values for higher-mass hadrons are substantially larger. Power-law exponent $n \approx 8.5$ for pions is also consistent with unidentified hadrons at 5 TeV and has a  $\log(\sqrt{s}/\text{10 GeV})$ energy dependence~\cite{alicetomspec}. Soft-component exponent $n$ values for more-massive hadrons are not well-defined because the hard-component contribution is much larger than for pions. Varying $n$ for $\hat S_0(y_t)$ then has little impact on the overall spectra.
  
 \begin{table}[h]
 	\caption{TCM model parameters for identified hadrons from 5 TeV \ppb\ collisions~\cite{ppbpid}. Numbers without uncertainties are adopted from a comparable hadron species with greater accuracy. Parameters $ \bar p_{tsi}$ and $\bar p_{th0i}$ are determined by model functions $\hat S_{0i}(y_t)$ and $\hat H_{0i}(y_t)$ with parameters from Table~\ref{pidparams}.  $h$ represents results for unidentified hadrons.
 	}
 	\label{otherparams}
 	\begin{center}
 		\begin{tabular}{|c|c|c|c|c|} \hline
 			&   $z_0$    &  $\tilde z_i$ &   $ \bar p_{ts}$ (GeV/c)  & $ \bar p_{th0}$ (GeV/c)  \\ \hline
 			$ h$        &  $\equiv 1$  & $\equiv 1$  & $ 0.40\pm0.02$ &    $1.30\pm0.03$  \\ \hline
 			$ \pi^\pm$        &   $0.70\pm0.02$  & $0.8\pm0.05$  & $0.40\pm0.02$ &    $1.15\pm0.03$  \\ \hline
 			$K^\pm $   &  $ 0.125\pm0.01$   &  $2.8\pm0.2$ &  $0.60$&  $1.34$   \\ \hline
 			$K_\text{S}^0$        &  $0.062\pm0.005$ &  $3.2\pm0.2$ &  $0.60\pm0.02$ &   $1.34\pm0.03$  \\ \hline
 			$p $        & $ 0.07\pm0.005$    &  $7.0\pm1$ &  $0.73\pm0.02$&   $1.57\pm0.03$   \\ \hline
 			$\Lambda $        &  $0.037\pm0.005$    & $7.0$ &   $0.76\pm0.02$ &    $1.65\pm0.03$ \\ \hline	
 		\end{tabular}
 	\end{center}
 \end{table}
  
Table~\ref{otherparams} shows fixed values of PID parameters $z_0$ and $\tilde z_i$ for five hadron species. The complete description of $z_{xi}(n_s)$ in Eq.~(\ref{rhosi}) with $\tilde z_i$ and $z_{0i}$ as derived from 5 TeV \ppb\ spectra is reported in Refs.~\cite{ppbpid,pidpart1,pidpart2}. Given the TCM expression in Eq.~(\ref{pidspectcm}) the correct rescale factor $1/\bar \rho_{si}$ derived from Eq.~(\ref{rhosi}) should result in \pbpb\ data spectra coincident with $\hat S_0(y_t)$ as $y_t \rightarrow 0$ for all centralities. 

\subsection{Terminology and data-model comparisons}

Application of the TCM to spectrum data requires certain distinctions between data structures and model elements. Equation~(\ref{rhotcm}) represents both model functions and the inferred structure of spectrum data based on empirical analysis of spectrum evolution with \nch\ and collision energy~\cite{ppprd,alicetomspec}. That process is not based on {\em a priori} assumptions about spectrum structure. Implicit in the analysis is the empirical determination that for \pp\ collisions $\bar \rho_h \approx \alpha \bar \rho_s^2$ with percent accuracy~\cite{ppprd}. Given spectrum structure suggested by Eq.~(\ref{rhotcm}), rescaled spectra $\bar \rho_0(y_t) / \bar \rho_s$ should have, in the limit $\bar \rho_{s} \rightarrow 0$, an asymptotic form represented by symbol $\hat S_0(y_t)$, a data ``soft component''  that can in turn be described by a simple model function also represented by symbol $\hat S_0(y_t)$. 

For hadron species $i$, subtracting model function $\hat S_{0i}(y_t)$ from data ratio $\bar \rho_{0i}(y_t) / \bar \rho_{si}$ leads to a data ``hard component'' represented by the expression  $\tilde z_i x(n_s) \nu(n_s) \hat H_{0i}(y_t)$. Rescaling that difference by product $\tilde z_i x(n_s) \nu(n_s)$ (factorization) leads to data component $\text{Y}_i(y_t)$ in Eq.~(\ref{pidhard}) that can in turn be approximated by a hard-component model function also represented by $\hat H_{0i}(y_t)$. Validity of that factorization is considered in Sec.~\ref{extend}. One could introduce different symbols for data  and model elements, but proliferation of symbols might be confusing. In the text a distinction is maintained between data and models.

\section{2.76 $\bf TeV$ $\bf Pb$-$\bf Pb$ collision geometry} \label{geometry}

2.76 TeV \pbpb\ and \pp\ PID spectrum data from Sec.~\ref{alicedata} are analyzed via a TCM developed for \ppb\ PID spectra in Ref.~\cite{ppbpid}. 
To interpret PID \pt\ spectra from Ref.~\cite{alicepbpbpidspec} properly geometry parameters should be estimated accurately, where ``geometry'' here refers to parameters such as $N_{part}$ and $N_{bin}$ but not centrality or impact parameter {\em per se}.
\pbpb\ collision geometry is determined first via conventional Glauber model (see Table~\ref{ppbparams1}). An alternative approach is based on \pbpb\ ensemble-mean \mmpt\ data as for the \ppb\ TCM analysis in Ref.~\cite{ppbpid}. Correction of proton spectrum inefficiencies is  addressed.

\subsection{Classical-Glauber Pb-Pb collision geometry} \label{geomparams}

The Glauber-model approach assumes that a number of participants $N_{part}$ and number of \nn\ binary collisions $N_{bin}$ may be inferred from a Glauber Monte Carlo model including certain assumptions about how \nn\ interactions may occur within \aa\ collisions. Results from a Glauber-model study of 2.76 TeV \pbpb\ collisions~\cite{pbpbcent} are summarized in Sec.~\ref{pbpbgeomx}. 
To implement the PID spectrum TCM in Sec.~\ref{pidspec} for 2.76 TeV \pbpb\ collisions based on the Glauber model an expression for $x(n_s)$ is required. For \pp\ collisions $x(n_s) \approx \alpha \bar \rho_s$ applies and $\nu \equiv 1$. For \ppb\ collisions $x(n_s)$ and $\nu(n_s)$ are derived from \mmpt\ data as reported in Ref.~\cite{tommpt}. For \aa\ collisions there is considerable uncertainty about $x(n_s)$ modeling.%
\footnote{The number of \nn\ binary collisions $N_{bin}$ derived from a Glauber MC applied to A-B collisions is suspect based on a comparison of TCM vs Glauber \ppb\ centrality trends reported in Ref.~\cite{tomglauber}.} 
However, within a Glauber context and based on the structure of Eq.~(\ref{pidspectcm}) $x(n_s)$ might be inferred from the measured  trend of midrapidity charge density $\bar \rho_0$ as
\bea \label{npartrho}
(2/N_{part}) \bar \rho_0 &=& \bar \rho_{sNN} [1 + x(n_s) \nu(n_s)].
\eea
Given charge density $\bar \rho_0$, Glauber MC estimates of $N_{part}$ and $\nu$ and an assumed {\em fixed value} of $\bar \rho_{sNN}$,  $x(n_s)$ can be derived from \pbpb\ data. A further assumption for that strategy is that there is no connection between $\bar \rho_{sNN}$ and $x$, whereas for \pp\ and \ppb\ spectra the direct connection $x \approx \alpha \bar \rho_{sNN}$ is {\em required} by spectrum and \mmpt\ data~\cite{ppprd,ppbpid,tomglauber}.

\begin{figure}[h]
	\includegraphics[width=3.3in]{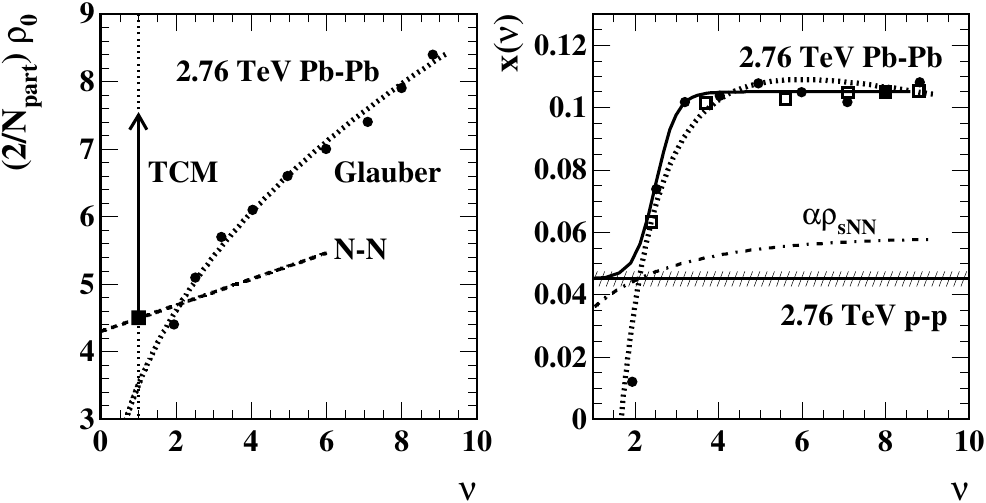}
	\caption{\label{900a}
		Left: Unidentified-hadron charge density $\bar \rho_0$ per participant-nucleon pair from Ref.~\cite{alicepbpbyields} (points) vs number of binary \nn\ collisions per participant pair $\nu$. 
The dashed curve corresponds to \nn\ linear superposition (fixed \pp\ value $x = 0.045$). The TCM vertical arrow and bold dotted curve are described in the text.
		Right: $x(\nu)$ vs $\nu$ transformed from the left panel (solid points, bold dotted) via inverting Eq.~(\ref{npartrho}) and (solid curve) as defined by Eq.~(\ref{16}). The dash-dotted curve is described in the text. Open boxes  represent Glauber $x(\nu)$ on $\bar \rho_0$ values for the present study. 
	}   
\end{figure}

Figure~\ref{900a} (left) shows quantity $(2/N_{part}) \bar \rho_0$ for 2.76 TeV \pbpb\ collisions (solid points) as reported in Ref.~\cite{alicepbpbyields}. Corresponding values of (possibly biased) $\nu'(n_s)$ are derived from Ref.~\cite{alicenonpidspec}.
The bold dotted curve is inferred from the Glauber power-law relations in Fig.~\ref{glauber} as $3.5 \nu^{0.40}$. The vertical arrow indicates a TCM trend inferred from \mmpt\ data with fixed values $\nu = N_{part}/2  =1$ for $\bar \rho_0 < 15$ as described in the next subsection.

Figure~\ref{900a} (right) shows an empirical expression for $x(n_s) \rightarrow x(\nu)$ (solid curve)
\bea \label{16}
x(\nu) &=& 0.045 + 0.06 \left\{1+ \tanh[2(\nu - 2.5)] \right\}/2,
\eea
where the constant term is $x_{pp} \equiv \alpha \bar \rho_{spp} = 0.0105 \times 4.3 \approx 0.045$ and the coefficient of the second term 0.060 is adjusted to describe the $\bar \rho_0$ data. The value $\bar \rho_{sNSD} = 4.3$ disagrees with the expectation $\bar \rho_{sNSD} \approx 0.81 \ln(\sqrt{s} / \text{10 GeV}) \approx 4.55$ for 2.76 TeV \pp~\cite{alicetomspec}. However, the $(2/N_{part}) \bar \rho_0$ data from Ref.~\cite{alicepbpbyields} require 4.3. 
The bold dotted curve is $x$ inferred by inverting Eq.~(\ref{npartrho}) assuming $\bar \rho_{sNN}$ is held fixed at 4.3 and with $(2/N_{part})\bar \rho_0$ given by the bold dotted curve at left.
The dash-dotted curve is Eq.~(\ref{npartrho}) inverted assuming $x = \alpha \rho_{sNN}$ (consistent with $x = \bar \rho_h/\bar \rho_s$ observed for \pp\ collisions) and that $\rho_{sNN}$ is the root of a quadratic equation (is {\em not} fixed).
Final Glauber values for $x(\nu)\nu$ from six centralities of \pbpb\ collisions are  0.152  0.375 0.575 0.744 0.839 0.926. Those values appear as open boxes in Fig.~\ref{inverse} (right).

\subsection{Alternative TCM  Pb-Pb collision geometry} \label{altgeom}

A study of Glauber centrality for 5 TeV \ppb\ collisions in Ref.~\cite{tomglauber} revealed  that a Monte Carlo method assuming the eikonal approximation returns much larger participant numbers than are consistent with \ppb\ \mmpt\ data. The large difference appears to result from {\em exclusivity}: A projectile nucleon can only interact with one target nucleon {\em at a time}, i.e.\ experiences a ``dead time'' before interacting again~\cite{tomexclude}. Introduction of an exclusivity constraint on a Glauber Monte Carlo returns numbers consistent with \mmpt\ data. A strategy based on \mmpt\ data is thus utilized here to estimate \pbpb\ geometry parameters.

Uncorrected (for low-\pt\ cutoff $p_{t,cut}$) \pbpb\ $\bar p_t'$ data have the same general TCM description as for \ppb\ data
\bea \label{ptprime}
\bar p_t' \equiv \frac{\bar P_t'} {n_{ch}'} &\approx & \frac{\bar P_{tsNN}' + \nu(n_s) \bar P_{thNN}(n_s)}{n_{sNN}'(n_s) + \nu(n_s) n_{hNN}(n_s)}~~~
\\ \nonumber
&\approx &  \frac{\bar p_{ts} + x(n_s)\, \nu(n_s) \bar p_{thNN}(n_s)}{\xi + x(n_s) \nu(n_s)},
\eea
where primes denote the effects of spectrum data integrated over a \pt\ interval with \pt\ acceptance lower limit $p_{t,cut} \approx 0.15$ GeV/c. The ratio of uncorrected to corrected soft multiplicity is $\xi = n_s' / n_s < 1$. Variation of integrated $\bar P_{ts}'$ with $p_{t,cut}$ is much less and $\bar P_{ts}' \approx \bar P_{ts} = n_s \bar p_{ts}$ is assumed. Uncorrected integrated charge is $n_{ch}' = \xi n_s + n_h$. The details of \pt\ production can thus be isolated by forming the product $(n_{ch}' / n_{s}) \, \bar p_t'$ to obtain
\bea \label{proddd}
\frac{n_{ch}' }{n_{s}}  \bar p_t'(n_s) &\approx& \frac{\bar P_t}{n_{s}} \approx \bar p_{ts} + x(n_s)   \,  \nu(n_s)\, \bar p_{thNN}(n_s).~
\eea
For the present study $\bar p_{thNN}(n_s)$ is assigned a fixed value.
Jet production is measured by increase of \mmpt\ above $\bar p_{ts}$.
Equation (\ref{proddd}) is crucial because $\bar \rho_s$ or $n_s$ cancels in ratio and product $x(n_s)\nu(n_s)$ can be inferred from \mmpt\ data.  

\begin{figure}[h]
	\includegraphics[width=1.65in]{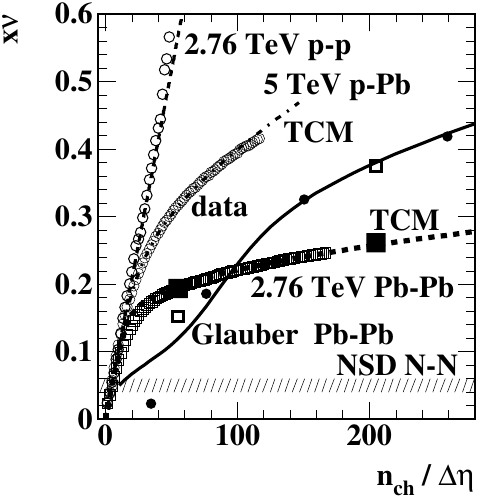}
	\includegraphics[width=1.65in]{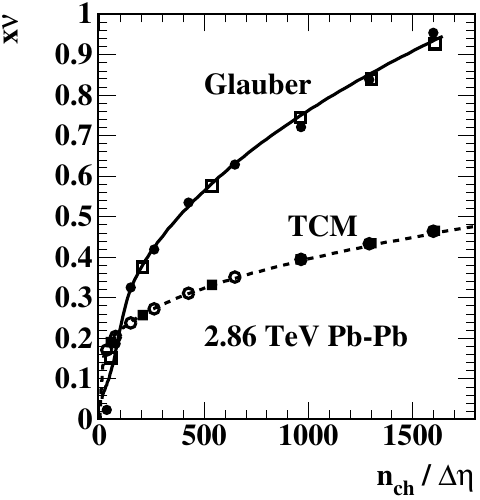}
	\caption{\label{inverse}
		Left: Product $x\nu$ (open points) extracted from \mmpt\ data via Eq.~(\ref{proddd}). Dash-dotted and upper dashed curve are TCM predictions. The lower dashed curve is a TCM description of \pbpb\ data derived from Eqs.~(\ref{xmodel}) and (\ref{nueq}). The solid points and curve are derived from Glauber solid points and curves in Fig.~\ref{900a}. 
		Right: Extrapolation of the TCM trend from the left panel over the full \pbpb\ multiplicity range. The Glauber points and curve are  from Fig.~\ref{900a} solid points and solid curves.
	} 
\end{figure}

Figure~\ref{inverse} (left) shows $x(n_s)\nu(n_s)$ (points) inferred by inverting \mmpt\ data from three collision systems: \pp, \ppb\ and \pbpb. Curves for \pp\ and \ppb\ systems are TCM descriptions. For instance,  $x(n_s) = \alpha \bar \rho_s$ ($\alpha \approx 0.0105$ for 2.76 TeV) and $\bar \rho_0 = \bar \rho_s + \alpha \bar \rho_s^2$  with {$\bar p_{thNN} \approx 1.20$ GeV/c} and $\nu = 1$ defines the dashed curve through 2.86 TeV \pp\ data. The dash-dotted curve for  \ppb\ data is obtained from Ref.~\cite{ppbpid} which presents a model for $x(n_s)$ derived from \mmpt\ data and expressed as
\bea \label{xmodel}
x(n_s) &=& \alpha \bar \rho_{sNN} =  \frac{\alpha}{\left\{[1/ \bar \rho_s]^{n_1} + [1/f(n_s)]^{n_1}\right\}^{1/n_1}},~~~
\eea
where $f(n_s) = \bar \rho_{s0} + m_0(\bar \rho_s - \bar \rho_{s0})$. The combination in Eq.~(\ref{xmodel}) ensures that the trend with lowest value determines $x(n_s)$, with smooth transition from \pn\ trend to \ppb\ trend determined by exponent $n_1 \approx 4$. Below a transition point at $\bar \rho_{s0} \approx 15$ $x(n_s) \approx \alpha \bar \rho_{sNN}$ as for \pp\ collisions. Above the transition $x(n_s)$ still increases linearly but with  reduced slope controlled by parameter $m_0 \approx 0.1$. Factor $\nu(n_s) = 2N_{bin}/N_{part}$ is obtained from the \pa\ relations $N_{part}/2 = \bar \rho_s / \bar \rho_{sNN} \approx \alpha \bar \rho_s / x$ and $N_{bin}  = N_{part} - 1$. Those trends reveal competition for jet-related \pt\ increase between individual \nn\ pairs measured by $x(n_s)$ and increasing \nn\ binary collisions measured by $\nu(n_s)$. \pbpb\ \mmpt\ results suggest that a similar competition occurs in \pbpb\ for more-peripheral collisions, but factorization of  $x\nu$ is not simple as  for \ppb.

For \pbpb\ data $m_0 \rightarrow 0$ is assumed as a trial estimate. Given the Glauber trends in Fig.~\ref{glauber} one might expect the approximate relation $\nu \propto \bar \rho_0^{0.38}$ for \pbpb\ particle densities above the transition point $\bar \rho_{s0}$. For smaller densities below the transition point one should expect $\nu \approx 1$ as for \pp. That description suggests a model for $\nu(\bar \rho_0)$ based on a variable power-law exponent
\bea
\mu(n_s) &=& 0.44 \{1+[\tanh((\bar \rho_s - 20)/15)-1]/2  \}
\eea
that transitions exponent $\mu$ from 0 below $\bar \rho_{s0}$ to 0.44 above  $\bar \rho_{s0}$. The expression for $\nu(n_s)$ is then
\bea \label{nueq}
\nu(n_s) &=& 1 + 0.25[(\bar \rho_s/13)^{\mu(n_s)} -1  ]
\eea
where that combination reduces the excursion of $\nu$ above 1 (in part with factor 0.25) and determines the other fixed values so as to accommodate the \pbpb\ \mmpt\ data as demonstrated by the dashed curve through \pbpb\ points.

Figure~\ref{inverse} (right) shows  Glauber model (solid dots and curve) and \pbpb\ \mmpt\ data (solid squares, dashed TCM extrapolation)  over the full range of \pbpb\ charge densities. The final TCM $x\nu$ values are 0.19, 0.26, 0.34, 0.40, 0.44 and 0.47 plotted as the solid squares in Fig.~\ref{inverse}. Those points indicate extrapolated model values corresponding to data $\bar \rho_0$ values, not separate measurements. These $x\nu$ estimates correspond to {\em integrated hard-component $P_t$} in Eq.~(\ref{ptprime}), not differential fragment distributions on \pt\ or \yt. Below is a more-detailed description of the model. 

\begin{figure}[h]
	\includegraphics[width=3.3in]{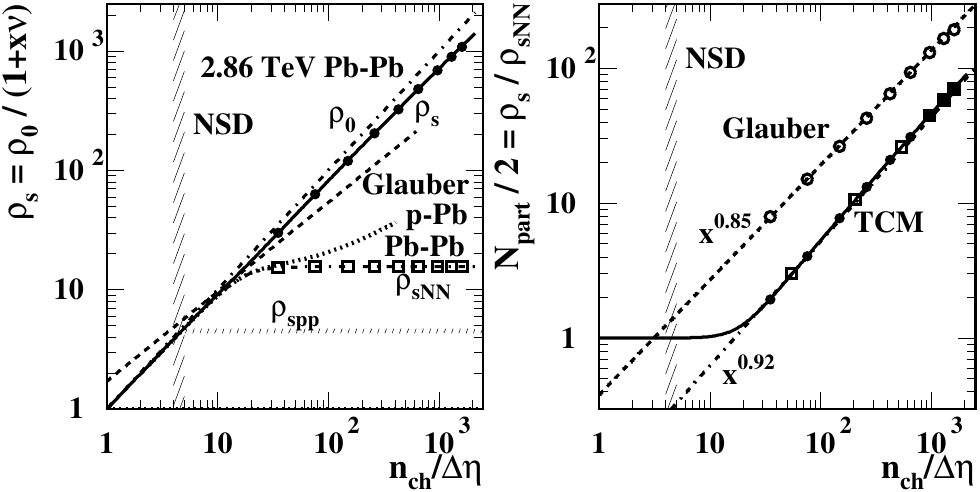}
\put(-137,80) {\bf (a)}
\put(-18,77) {\bf (b)} \\
	\includegraphics[width=3.3in]{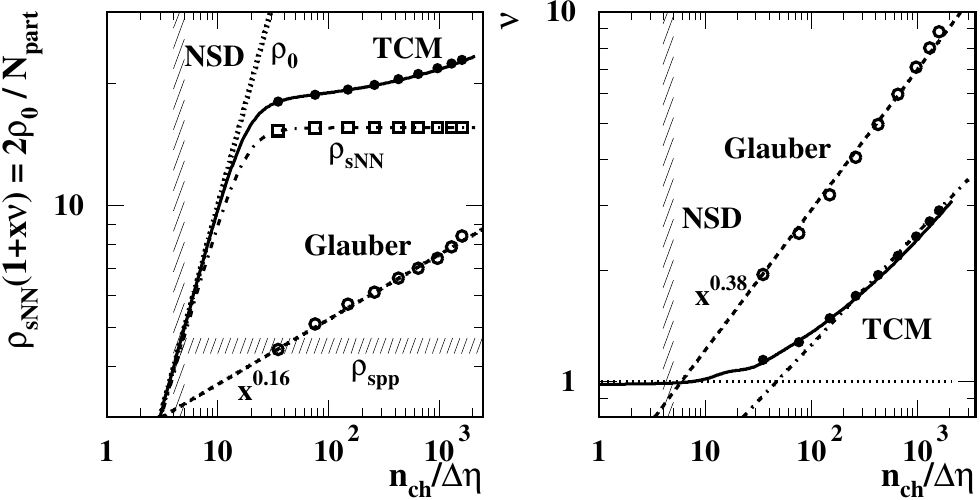}
\put(-138,94) {\bf (c)}
\put(-20,90) {\bf (d)} 
	\caption{\label{story}
		(a) Soft density $\bar \rho_s = \bar \rho_0 / (1+x\nu)$ (solid dots and curve) obtained with TCM $x\nu$ from Fig.~\ref{inverse}. The dash-dotted curve is a $\bar \rho_0$ reference. The dashed curve is obtained with Glauber $x\nu$ in  Fig.~\ref{inverse}. The lower dash-dotted and dotted curves are TCM $\bar \rho_{sNN}$ trends for \pbpb\ and \ppb\ collisions respectively.
		(b) $N_{part}/2 = \bar \rho_s / \bar \rho_{sNN}$ (TCM points and curve) derived from $\bar \rho_s$ (solid) and $x = \alpha  \bar \rho_{sNN}$ conjecture (lower dash-dotted) in (a). Open circles are Glauber (primed) values from Table~\ref{ppbparams1}.
		(c) $2 \bar \rho_0 / N_{part} = \bar \rho_{sNN} (1+x\nu)$ for TCM (solid points and curve) and Glauber model (open points and dashed curve).
		(d) Quantity $\nu = 2 N_{bin}/N_{part}$ from the TCM (solid points and curve) and from the Glauber model (open circles). The dashed power-law curve is derived from Fig.~\ref{glauber}.
	} 
\end{figure}

Figure~\ref{story} (a) shows soft-component density $\bar \rho_s$ (points, solid curve) derived from measured density values $\bar \rho_0$ from Ref.~\cite{alicepbpbyields} (solid points) as $\bar \rho_s = \bar \rho_0/(1+x\nu)$ where product $x\nu$ is inferred from \mmpt\ data in Fig.~\ref{inverse} (solid squares, dashed curves). The upper dash-dotted line shows $\bar \rho_0 = n_{ch} / \Delta \eta$ for comparison. The dashed line shows  $\bar \rho_s$ estimated with product $x\nu$ from the Glauber system. See Fig.~\ref{inverse} (right) for $x\nu$ comparisons. The lower dash-dotted and dotted curves are $\bar \rho_{sNN}$ for \pbpb\ and \ppb\ data. The open boxes are $\bar \rho_{sNN}$ values for \pbpb\ spectrum analysis. 

Figure~\ref{story} (b) shows TCM estimates for $N_{part}/2 = \bar \rho_s / \bar \rho_{sNN}$ (solid points, solid curve) based on estimates of $\bar \rho_{sNN}$ shown in panel (a), (lower dash-dotted). Open circles are  Glauber estimates vs $\bar \rho_0$ from Fig.~\ref{glauber} (right).

Figure~\ref{story} (c) shows TCM trend $2\bar \rho_0 / N_{part} = \bar \rho_{sNN} (1 + x\nu)$ vs $\bar \rho_0$ (solid dots and curve). Also shown is the result from Glauber analysis (open circles), i.e.\  solid dots in Fig.~\ref{900a} (left) from Ref.~\cite{alicepbpbyields}. Open circles from the Glauber treatment present a dramatic contrast to  solid dots via TCM. The dashed curve is a power law $2.5\bar \rho_0^{0.16}$ inferred from trends shown in Fig.~\ref{glauber}. The dash-dotted curve is an estimated trend for quantity $\bar \rho_{sNN} = x(n_s) / \alpha$ with $x(n_s)$ defined by Eq.~(\ref{xmodel}). In effect, $\bar \rho_{sNN} = \bar \rho_s$ ($\nu = 1$) up to the transition point $\bar \rho_{s0} \approx \bar \rho_0 \approx 15$ and then remains approximately constant for more-central collisions.

Figure~\ref{story} (d) shows quantity $\nu$ (solid dots and curve) derived from product $x\nu$ inferred from \mmpt\ data and the \pbpb\ $\bar \rho_{sNN} \leftrightarrow x(n_s)/\alpha$ hypothesis shown in panel (a). Also shown are $\nu'$ values (open circles) derived from Glauber analysis  in Ref.~\cite{pbpbcent} that follow a power-law trend $\propto \bar \rho_0^{0.38}$ (dashed) consistent with trends in Fig.~\ref{glauber}. The \mmpt\ TCM data trend approaches the full Glauber power-law {\em variation} but {\em three times lower in magnitude}.

The trends in Fig.~\ref{story} are explained as follows: For peripheral collisions the three collision systems share a common \nn\ straight-line trend $x \approx \alpha \bar \rho_0$ with $\nu \approx 1$ corresponding to the congruences in Fig.~\ref{inverse} (left). \ppb\ and \pbpb\ collisions are restricted to single peripheral \nn\ collisions due to {\em exclusivity within a small overlap volume}. Then \ppb\ and \pbpb\ trends deviate from that linear trend above a transition near $\bar \rho_s = 15$. The more-central trend for \ppb\ data can be accurately predicted because $N_{part} = N_{bin} + 1$ and all geometry parameters are therefore dependent on the single parameter $x(n_s)$ that can be simply modeled as in Ref.~\cite{ppbpid}. The basis for a \pbpb\ model is discussed above in relation to Eq.~(\ref{nueq}).

The Glauber trends are dramatically different. Based on Fig.~\ref{900a} and Eq.~(\ref{npartrho}), for peripheral collisions $x \approx \alpha \bar \rho_{sNSD} \approx 0.05$, i.e.\ constant, and $x\nu$ is therefore $\propto \nu$ in contrast to the \mmpt\ data trend $x \propto \bar \rho_s$ and $\nu \approx 1$. Referring to Fig.~\ref{inverse}, above a transition near $\bar \rho_0 = 100$ $x$ is again approximately constant (0.105) and $x\nu$ is again $\propto \nu$, reaching values for central \pbpb\ twice the extrapolation (dashed) inferred by inverting \mmpt\ data. The large difference between Glauber results and \mmpt\ data arises from assumptions that $\bar \rho_{sNN}$ (and therefore $x$) is fixed at lower multiplicities, that $N_{part} \propto \bar \rho_0$ (see Fig.~\ref{glauber}), that nucleons may interact simultaneously with multiple nucleons and that  the Glauber model provides accurate estimates of \aa\ participant $N_{part}$ and binary-collision $N_{bin}$. 

Those assumptions are strongly contradicted by data. Above the transition near $\bar \rho_s \approx 15$ $N_{part}$ (and $\nu$) does increase, but the increase is strongly reduced from Glauber expectations as illustrated by factor 0.25 in the expression for $\nu_\text{PbPb}$ above, implying that Glauber-predicted binary collisions are reduced {\em three-to-four-fold} by exclusivity (again similar to the well-defined case for \ppb\ collisions). The assumption that $x$ remains fixed above $\bar \rho_{s0}$ might be questioned, but if $x$ does increase  exclusivity suppression of $\nu$ must be {\em even greater} according to data.

The dramatic difference in numerical values is of interest. For most-central \pbpb\ collisions $N_{part} = 383$ for Glauber (Table~\ref{ppbparams1}) vs 140 for TCM, and $N_{bin} = 1685$ for Glauber vs 210 for TCM. The average number of binary collisions per nucleon pair is $\nu = 9$ for Glauber and 3 for TCM. What seems a high-density system with many \nn\ collisions according to Glauber is, via \mmpt\ data, a low-density system with many fewer \nn\ collisions, including a substantial peripheral interval where only a single \nn\ collision for each participant is allowed by exclusivity. Also note that the number of {\em spectators} per nucleus (nucleons not interacting) for central \pbpb\ collisions is $208 - 70 = 138$ for the TCM geometry vs $208 - 192 = 16$ for the Glauber Monte Carlo. The main difference is lack of collision partners for many nucleons due to exclusivity.

For  \pbpb\ spectrum analysis values of $\bar \rho_s$ and $\nu$ (unprimed values) are shown in Table~\ref{ppbparams1} with $x =  \alpha \bar \rho_{sNN}= 0.163$ for all six centrality classes. PID spectra from two event classes (mid-central class 4 and peripheral class 7) of 5 TeV \ppb\ collisions as reported in Ref.~\cite{ppbpid} and presented in Table~\ref{rppbdata} (unprimed values) are included in this analysis as references. For class 4 $x = 0.16$, $\nu = 1.26$, $ \rho_{sNN} = 14.1$ and $N_{part}/2 = 1.35$. For  class 7 $x = 0.048$, $\nu \equiv 1$, $ \rho_{sNN} = 4.2$ and $N_{part}/2 \equiv 1$. Note that $\rho_{sNN} = 4.2$ is significantly below the NSD \pp\ value $\approx 5$~\cite{alicetomspec} suggesting that jet production may be suppressed for that  class.
2.76 TeV \pp\ spectra are treated as special cases of \pbpb\ spectra with $N_{part}/2 = N_{bin} = 1$.

\subsection{Correcting Pb-Pb and p-p proton inefficiencies} \label{correct}

As reported in Refs.~\cite{pidpart1,pidpart2} all 5 TeV \ppb\ {\em neutral} $K^0_\text{S}$ and Lambda data agree with the PID spectrum TCM within data uncertainties. In contrast, hard components for identified {\em charged} hadrons based on $dE/dx$ measurements deviate  from TCM predictions: substantial excess for charged pions and strong suppression for protons. Further indication of a problem is provided by 2.76 TeV \pp\ PID spectra that, for pions and charged kaons, approximate the \ppb\ results. Proton spectra from \pp\ and \pbpb\ collisions thus appear problematic as discussed below. In Sec.~III B of Ref.~\cite{pidpart1} a procedure is described to correct proton data for $\approx 40$\% suppression. Pion data are of insufficient quality to correct a corresponding $\approx 10$\% excess due to apparent pion-proton misidentification.

\begin{figure}[h]
	\includegraphics[width=1.65in]{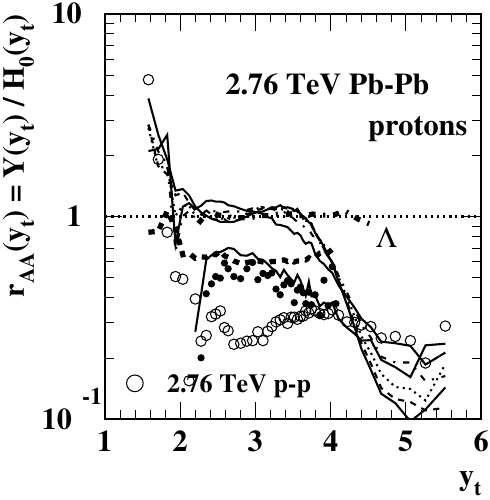}
	\includegraphics[width=1.65in]{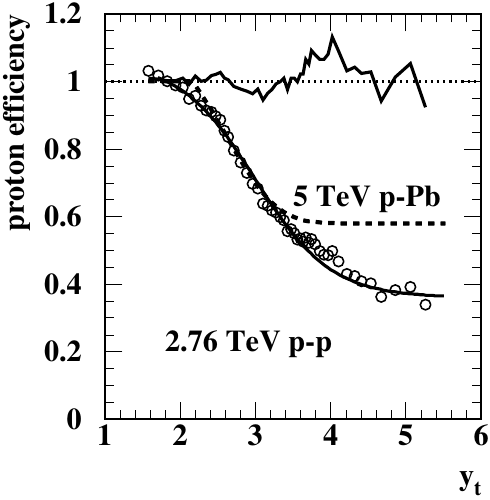}
	\caption{\label{rcpz}
		Left: \pbpb\ proton hard-component ratios $r_\text{AA}(y_t)$ as in Fig.~\ref{protons} (d) but uncorrected for inefficiencies. Bold dash-dotted is \ppb\ Lambdas and bold dashed is uncorrected \ppb\ protons.
		Right: Proton efficiencies for 2.76 TeV \pp\ (present study) and 5 TeV \ppb\ (see Ref.~\cite{pidpart1}) spectra. 
	}   
\end{figure}

Figure~\ref{rcpz} (left) shows $r_\text{AA}$ ratios from Fig.~\ref{protons} (d) before correction. The \ppb\ Lambda data (dash-dotted) {\em follow the TCM prediction} within data uncertainties whereas corresponding proton data (dashed) show the 40\% suppression. 2.76 TeV \pp\ data (open circles) show greater suppression plus irregularities. 
The procedure of Ref.~\cite{pidpart1} is applied to \pp\ data (assuming that collision system is simple) to derive a correction for \pbpb\ proton spectra.

Figure~\ref{rcpz} (right) shows the 2.76 TeV \pp\ proton spectrum in ratio to the TCM prediction (open circles). The systematic fall-off at high \yt\ common to three hadron species  in Figs.~\ref{pions} (d), \ref{kaons} (d) and \ref{protons} (d) (e.g.\ bold solid curve in Fig.~\ref{protons} (d)) should not be corrected and is removed here by the tanh function described for Fig.~\ref{pions} (d). 
A tanh function showing the correction trend (lower solid curve) $0.77(1-\tanh(y_t-2.9))/2+0.31$ provides efficiency corrections. The upper solid curve is the corrected \pp\ spectrum ratio. The dashed curve shows the correction for 5 TeV \ppb\ proton spectra as reported in Ref.~\cite{ppbnmf}.

It could be argued that such a correction is arbitrary, based on a TCM described as ``just fitted to the data.'' However, detailed study of PID spectrum systematics in Refs.~\cite{ppbpid,pidpart1,pidpart2} reveals that each hadron species is transported from soft to hard component (via jet production) with {\em conservation of species particle number} as described in Ref.~\cite{transport}. Thus, knowledge of a soft-component density below 0.5 GeV/c enables quantitative prediction of a hard-component density above 5 GeV/c. Species not relying on $dE/dx$ PID are consistent with such predictions (e.g.\ \ppb\ $K_\text{S}^0$ and Lambdas  in Figs.~\ref{kaons} and \ref{protons} (d) below. Also see results for Cascades and Omegas in Ref.~\cite{ppsss}.

\section{2.76 $\bf TeV$ $\bf Pb$-$\bf Pb$ PID spectrum TCM} \label{pidspec}

2.76 TeV \pbpb\ and \pp\ PID spectrum data from Sec.~\ref{alicedata} are analyzed via a TCM developed for \ppb\ PID spectra in Ref.~\cite{ppbpid}.  
Instead of the  conventional Glauber model (see Table~\ref{ppbparams1})
an alternative approach in Sec.~\ref{altgeom} is based on \pbpb\ ensemble-mean \mmpt\ data as for the \ppb\ analysis in Ref.~\cite{ppbpid}. \pbpb\ spectrum hard components for three hadron species are isolated and compared to hard components for \ppb\ and \pp\ collisions as references. 80-90\% central 2.76 TeV \pbpb\ data from  Ref.~\cite{alicepbpbspecx} are also included as a reference, although the \pt\ acceptance is limited. Evolution of rescaled \pbpb\ spectra in the form $\text{X}_i(y_t)$ in relation to the \pp\ reference spectrum suggests {\em a new form of analysis with surprising implications}.

\subsection{$\bf Pb$-$\bf Pb$  PID spectrum plotting formats} \label{piddiff}

In spectrum figures below, published 2.76 TeV \pbpb\ and \pp\ PID \pt\ spectra introduced in Sec.~\ref{alicedata} are replotted in panels (a) as densities on \pt\ multiplied by $2\pi$ (consistent with $\eta$ densities from other TCM analyses and no rescaling by powers of 2).  \pbpb\ data are plotted as solid curves and  \pp\ data are plotted as open circles. 
In this unrescaled format the large gap in \pbpb\ event classes for peripheral collisions is apparent as well as close spacing of  the most-central three event classes. Consequences of that distribution are discussed in Secs.~\ref{revisit} and \ref{complement}.

The advantage of spectrum plots on transverse rapidity \yt\ should be apparent: At lower \yt\ spectra are slowly varying, permitting precise differential  comparisons. At higher \yt\ spectra are well approximated by linear trends corresponding to a power-law on \pt\  in a log-log format which also provides precise differential comparisons.

Panels (b) show rescaled \pbpb\ spectra in the form $\text{X}_i(y_t)$ defined by Eq.~(\ref{pidspectcm}) (thin solid) compared to TCM soft components $\hat S_{0i}(y_t)$ (dotted). In that format relative spectrum normalizations can be examined at low \yt\ at the percent level. Also plotted are rescaled \pp\ spectra (open circles).

Panels (c) show \pbpb\ spectrum hard components in the form $\text{Y}_i(y_t)$ defined by Eq.~(\ref{pidhard}) (thin solid) compared to hard-component models $\hat H_{0ipp}(y_t)$ (dotted).  Also plotted are hard components from \pp\ spectra (circles). 

Panels (d) show hard-component ratios $r_{\text{AA}i}(y_t) \equiv \text{Y}_i(y_t) / \hat H_{0ppi}(y_t)$ that can be related to conventional spectrum ratios $R_{\text{AA}i}(p_t)$ defined by Eq.~(\ref{raa}) and appearing in Fig.~\ref{piddata} (right panels). Whereas conventional $R_{\text{AA}i}(p_t)$ (below $p_t \approx 4$ GeV/c) approaches a reference trend increasingly insensitive to jet structure (thin solid reference curves in Fig.~\ref{piddata} (right)) ratio $r_{\text{AA}i}(y_t)$ (representing {\em all} jet fragments) is sensitive to jet structure to  below 0.5 GeV/c ($y_t \approx 2$), thereby including the mode of the hard component near $y_t \approx 2.7$ ($p_t \approx 1$ GeV/c).

\subsection{Differential $\bf Pb$-$\bf Pb$  PID spectrum data} \label{piddiff2}

\begin{figure}[h]
	\includegraphics[width=3.3in]{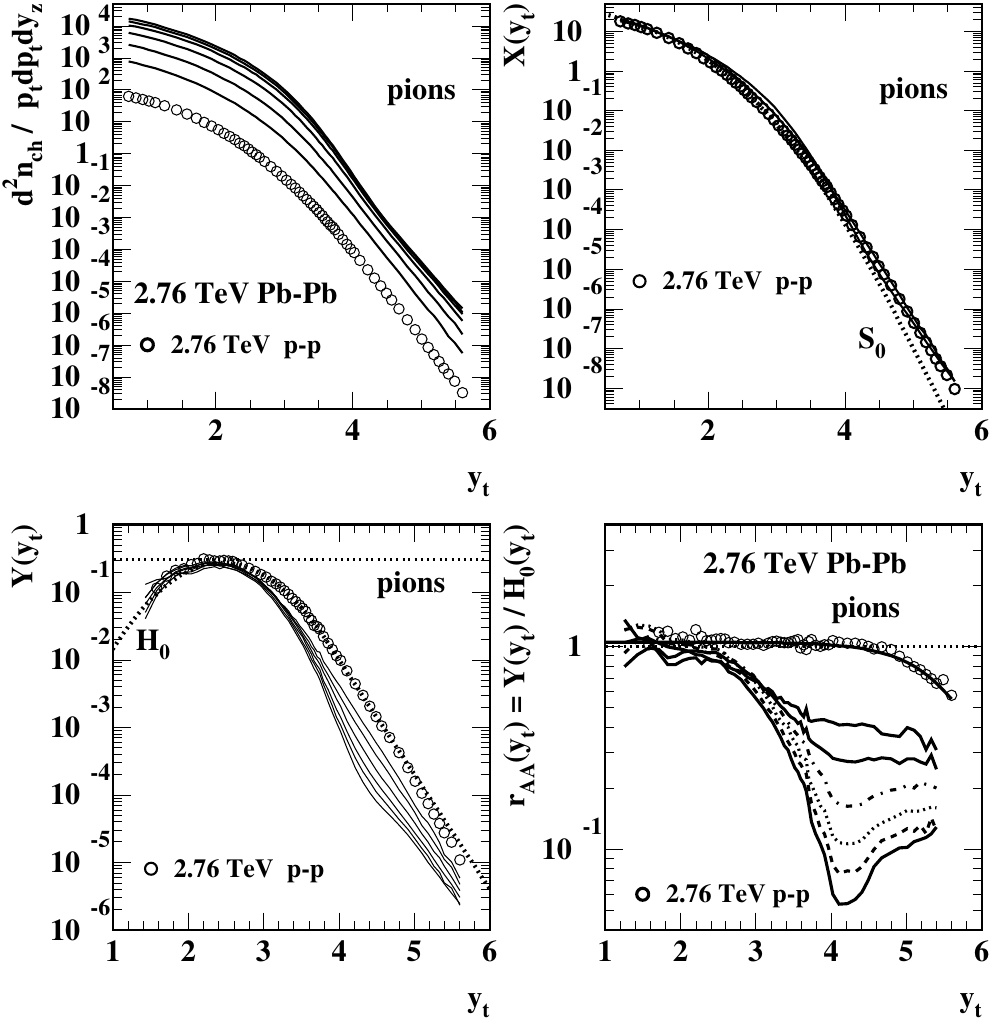}
\put(-142,230) {\bf (a)}
\put(-25,230) {\bf (b)}
\put(-142,90) {\bf (c)}
\put(-19,92) {\bf (d)}
	\caption{\label{pions}
		(a) Pion spectra from six event classes of 2.76 TeV \pbpb\ collisions as published (solid curves) and \pp\ spectra (open circles)~\cite{alicepbpbpidspec}  plotted on pion transverse rapidity $y_{t\pi}$. 
		(b) Data from panel (a) plotted in format $\text{X}(y_t)$ defined by the third line of Eq.~(\ref{pidspectcm}). 
		(c)  Data from panel (b) plotted in format $\text{Y}(y_t)$ defined by Eq.~(\ref{pidhard}).
		(d)   Data from (c) plotted as hard-component ratios $r_\text{AA}(y_t)$ defined by Eq.~(\ref{raax}).
	}  
\end{figure}
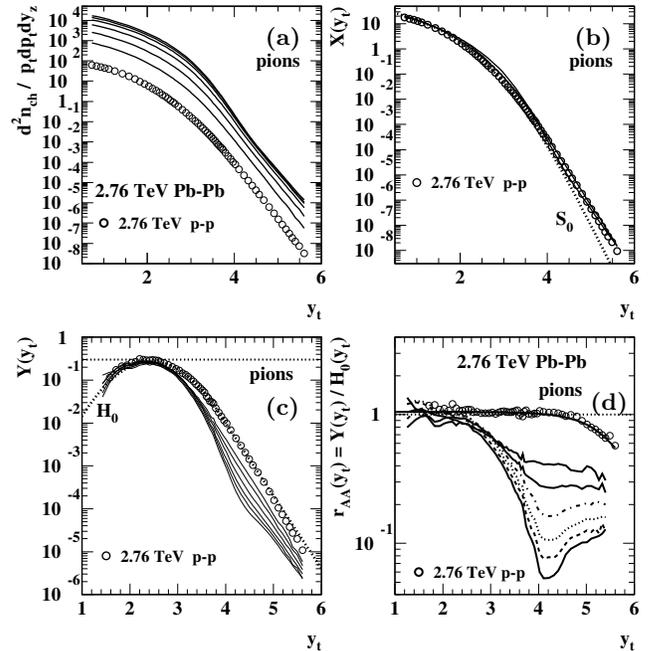

Figure~\ref{pions} shows data for identified pions from 2.76 TeV \pbpb\ (solid) and \pp\ (open circles).
The \pbpb\ pion data extend down to $p_t = 0.1$ GeV/c  ($y_t \approx 0.7$). 
Although panel (b) presents full spectra as differentially as possible systematic trends for pion spectra are difficult to discern albeit still visible. Such details are inaccessible in the plot format of Fig.~\ref{piddata} (a). 

Figure~\ref{pions} (c) shows finer details. The data here are transformed to densities on \yt\ compatible with the simple shape of model $\hat H_{0i}(y_t)$ in that format. The dotted line at 0.3 in this and other panels (c) is a reference to check for correct normalization of model $\hat H_0(y_t)$; the maxima vary relative to that reference depending on peak widths. 

Figure~\ref{pions} (d) shows hard-component ratios $r_{\text{AA}i}(y_t)$. 
The hard-component \pp\ data trend descends below the usual $\hat H_0(y_t)$ model at higher \pt\ (e.g.\ above 10 GeV/c). The solid curve through \pp\ points is a model $A(1-\tanh((y_t - 5.65)/0.95))/2$, {\em independent of hadron species}, required in Sec.~\ref{correct} for efficiency corrections.

\begin{figure}[h]
	\includegraphics[width=3.3in]{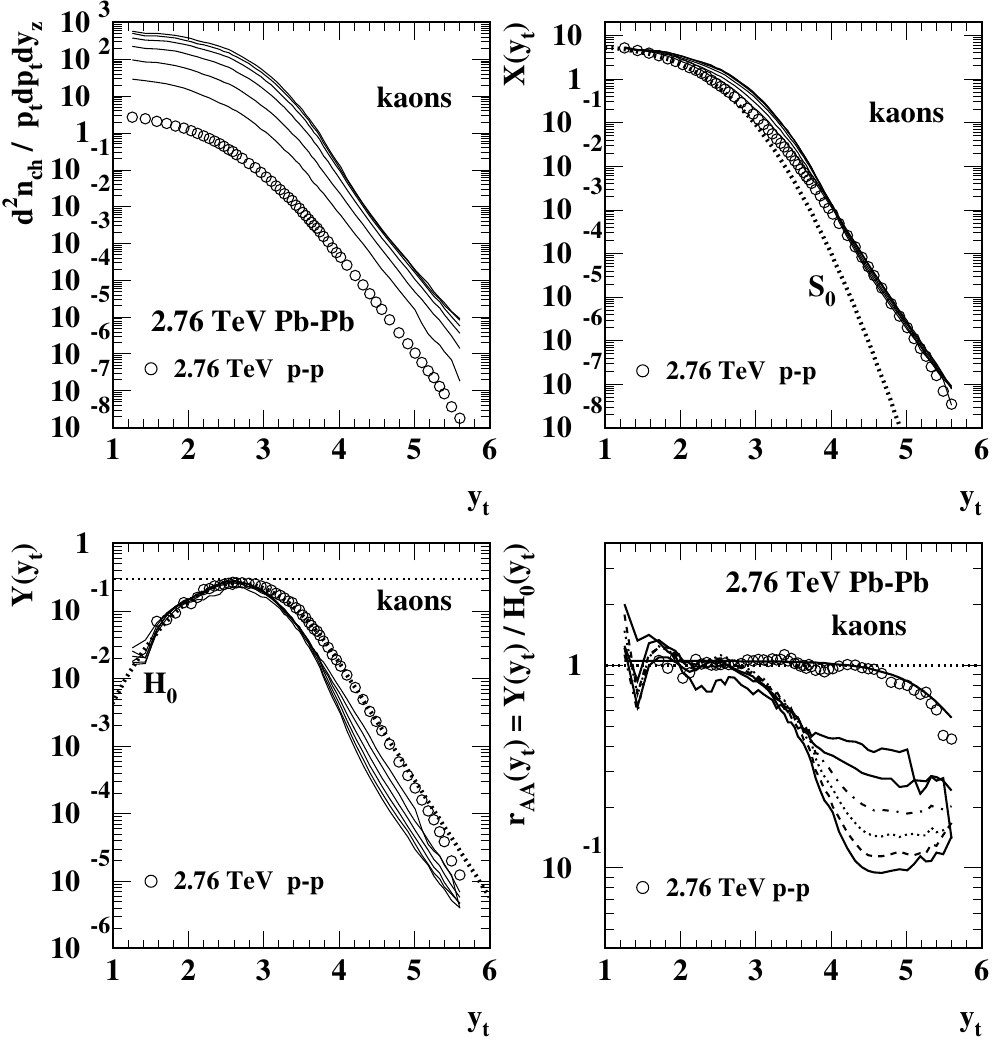}
\put(-142,231) {\bf (a)}
\put(-25,231) {\bf (b)}
\put(-142,92) {\bf (c)}
\put(-19,92) {\bf (d)}
	\caption{\label{kaons}
Kaon spectra from 2.76 TeV \pbpb\ and \pp\ collisions ($K^\pm$)
plotted as in Fig.~\ref{pions}. 
	}   
\end{figure}
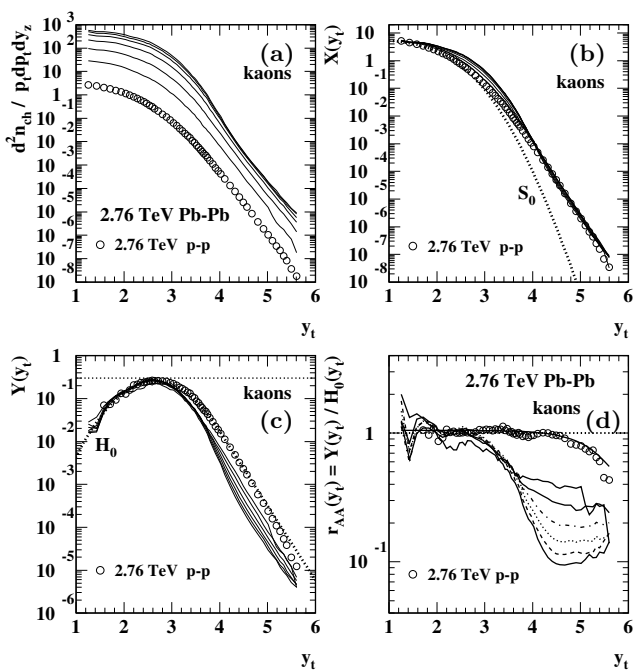

Figure~\ref{kaons} shows results for \pbpb\ charged kaons similar to those for pions. The most obvious difference is the larger fractional contribution from jet fragments near $y_t = 3$ relative to the kaon soft component, especially for more-central \pbpb\ collisions where the hard component remains dominant down to \yt\ = 2 ($p_t \approx 0.5$ GeV/c). 

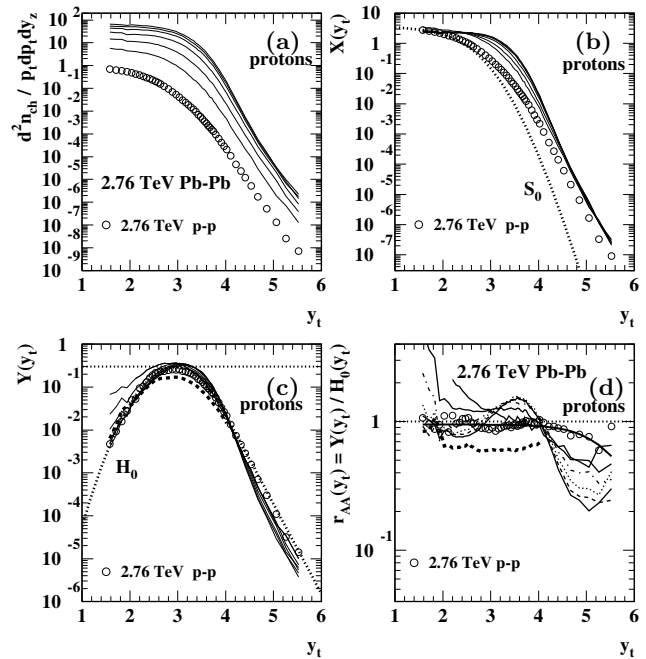
\begin{figure}[h]
	\includegraphics[width=3.3in]{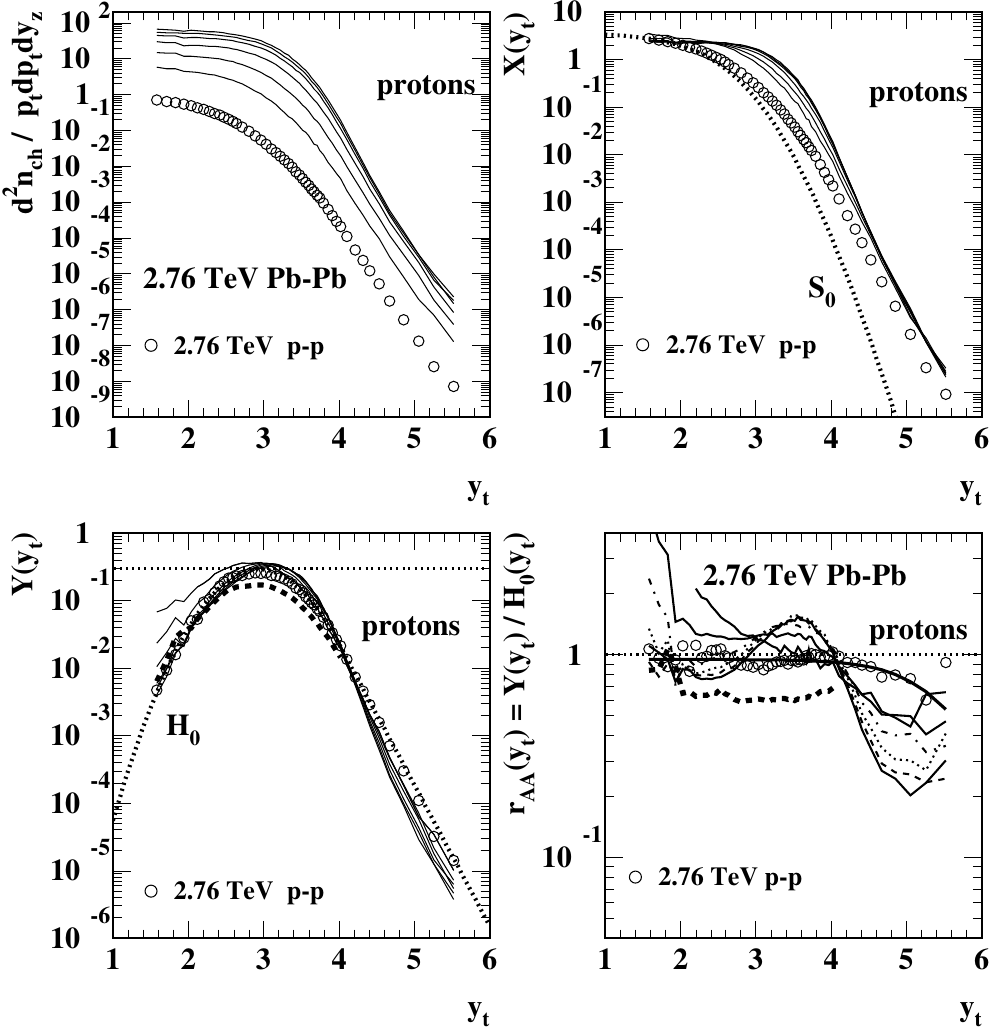}
	\put(-142,231) {\bf (a)}
	\put(-25,231) {\bf (b)}
	\put(-142,100) {\bf (c)}
	\put(-20,100) {\bf (d)}
	\caption{\label{protons}
Proton spectra from 2.76 TeV \pbpb\ and \pp\ collisions ($p,\,\bar p$)
as in Fig.~\ref{pions}. 5 TeV \ppb\ proton spectra for event class 4 (dashed), included as a reference in (c,d), are uncorrected and are suppressed relative to  TCM by about 40\%. 
	}  
\end{figure}

Figure~\ref{protons} shows results for protons from 2.76 TeV \pbpb\ (solid curves) and \pp\ (open circles) collisions. 
Also shown in panels (c,d) are 5 TeV \ppb\ results for protons (dashed) from event class 4 (mid-central). 
Data for \ppb\ protons are {\em un}corrected for large inefficiencies at higher \pt\ discussed in Sec.~\ref{correct}. Inefficiency of  \ppb\ protons 
is evident. Deviation (suppression) of \pbpb\ protons at higher \yt\ is relatively much less  than  for pions and kaons, as indicated by  comparing $R_\text{AA}$ data to thin {\em reference curves} in Fig.~\ref{piddata}.

\subsection{2.76 $\bf TeV$ $\bf p$-$\bf p$ and $\bf Pb$-$\bf Pb$ TCM parameters} \label{pidparamsx}

The PID spectrum TCM as expressed in Eqs.~(\ref{pidspectcm}) and (\ref{rhosi}) requires accurate parameter values for each collision system and hadron species. Previous results are presented in Tables~\ref{pidparams} and \ref{otherparams} from analysis of 5 TeV \ppb\ PID spectrum data in Ref.~\cite{ppbpid}. Parameters for TCM model functions $\hat S_{0i}(y_t)$ and $\hat H_{0i}(y_t)$ and PID parameters are itemized below for each hadron species. TCM parameters are required to be consistent across A-B systems and collision energies and are not adjusted to accommodate specific spectrum data.

For pions, the soft-component parameters are as in Table~\ref{pidparams} except for \pp\ where $T \rightarrow 140$ MeV. Hard-component parameters differ somewhat from the \ppb\ values in Table~\ref{pidparams}, with $\bar y_t = 2.45$, $\sigma_{y_t} = 0.60$ and $q = 3.9$. Only the $\bar y_t$ variation is significant relative to uncertainties in Table~\ref{pidparams}. Parameters $z_0$ and $\tilde z$ are as in Table~\ref{otherparams} except $z_0 \rightarrow 0.8$ for \pbpb. For pions $\hat S_{0i}(y_t)$ must be supplemented to include a resonance contribution as described in Sec. III A of Ref.~\cite{pidpart1}. The same resonance description is applied to all collision systems. 

For kaons,  soft-component parameters are as in Table~\ref{pidparams} except for \pp\ where $T \rightarrow 190$ MeV.  Hard-component parameters are as in Table~\ref{pidparams}.  Parameters $z_0$ and $\tilde z$ are as in Table~\ref{otherparams} except $\tilde z \rightarrow 2.8$ for $K^0_\text{S}$ data (within estimated uncertainties).
For protons and Lambdas  soft-component model parameters are as in Table~\ref{pidparams} and hard-component parameters are as in Table~\ref{pidparams}. Parameters $z_0$ and $\tilde z$ are as in Table~\ref{otherparams}. 

As a general conclusion there appear to be no substantial differences required among detailed {\em reference} spectrum models $\hat S_{0i}(y_t)$ and $\hat H_{0i}(y_t)$ for \pp, \ppb\ and \pbpb\ spectra. {\em Apparent} evidence for \pbpb\ jet modification in $r_\text{AAi}(y_t)$ appears only {\em above} \yt\ = 3.5 ($p_t \approx $ 2.3 GeV/c).

\section{Revisiting the PID spectrum TCM} \label{extend}

A notable feature of rescaled spectra $\text{X}_i(y_t)$ defined in Eq.~(\ref{pidspectcm}) and appearing in panels (b) of Figs.~\ref{pions}, \ref{kaons} and \ref{protons} is a {\em precise congruence} between \pbpb\ data spectra (curves) for all event classes and \pp\ data (open circles) above $y_t = 4$ (for mesons) or 4.5 (for protons). A similar feature for 200 GeV \auau\ spectra was observed for pions and protons in Ref.~\cite{hardspec}. It seems unlikely that jet suppression by a dense medium, supposedly varying strongly with particle density and therefore event class, would lead to such uniformity. Another anomaly is the {\em lack of suppression} of hard components at {\em lower} \pt\ in panels (c) above. A dense medium might be expected to fully suppress (thermalize?) fragments from jets with energies as low as 3 GeV~\cite{minijets}. Those unexpected results suggest reconsidering the TCM analysis procedure.

The expression for $\text{X}_i(y_t)$ relies only on determination of product $x\nu$ empirically from \mmpt\ data and hard/soft abundance ratio $\tilde z_i$ that again empirically depends only on hadron mass. One then has $\bar \rho_{si} = \bar \rho_{0i}/(1+ \tilde z_i x\nu)$ and $\text{X}_i(y_t) = \bar \rho_{0i}(y_t) / \bar \rho_{si}$. However the next step, determination of data hard component $\text{Y}_i(y_t)$, depends not only on forming the difference $\text{X}_i(y_t) - \hat S_{0i}(y_t)$ based on empirically-determined model $\hat S_{0i}(y_t)$ but also on division by product $\tilde z_i x\nu$ in Eq.~(\ref{pidhard}), which in turn relies on the assumption that $ \tilde z_i(n_s) x(n_s)\nu(n_s) \hat H_{0i}(y_t,n_s)$ factorizes and that a {\em data} hard component corresponding to model function $\hat H_{0i}(y_t,n_s)$ can thereby be isolated. That critical factorization assumption is reconsidered here.

Given soft-rescaled spectra for collision system XY (\pp, \ppb\ or \pbpb) denoted here by $\text{X}_{\text{XY}i}(y_t)$ 
define a modified form of hard-component measure $\text{Y}_i(y_t)$
\bea \label{yprime}
\text{Y}'_{\text{XY}i}(y_t) &=& \text{X}_{\text{XY}i}(y_t) - \hat S_{0i}
\eea
{\em without} hard rescaling by $\tilde z_i x\nu$ per Eq.~(\ref{pidhard}). $\text{Y}'_{\text{XY}i}(y_t)$ is the data equivalent of product $\tilde z_i(n_s) x(n_s)\nu(n_s) \hat H_0(y_t)$ and {\em remains a density on \pt}. Then define spectrum ratio
\bea \label{ry}
r_{\text{Y}i} &=& \text{Y}'_{\text{XY}i}  /\text{Y}'_{\text{NN}i} 
\\ \nonumber
&\approx & \frac{\tilde z_i x_{\text{XY}}\nu_{\text{XY}} \hat H_{0i\text{XY}}}{\tilde z_ix_\text{NN}\hat H_{0i\text{NN}}}
\eea
that is analogous to TCM ratio $r_\text{AA}$ in panels (d) of  Figs.~\ref{pions}, \ref{kaons} and \ref{protons} but in this case includes no assumption about XY hard-component factorization. This method is applied below to \pbpb\ and \ppb\ collision systems. $x_\text{NN} \approx 0.06$ is held fixed for three hadron species and $\tilde z_i \approx$ 0.9, 3 and 6 for pions, charged kaons and protons from Fig.~\ref{tildezparams} (left). Note that the numerator of Eq.~(\ref{ry}) represents intact hard components of XY data spectra whereas the denominator is a reference composed of products of {\em measured} $\tilde z_i$ and $x_\text{NN}$ and hard-component models $\hat H_{0i\text{NN}}(y_t)$ obtained from previous \pp\ and \ppb\ analysis. $\hat H_{0i\text{NN}}$ is in  this case transformed to a density on \pt.

\subsection{2.76 TeV Pb-Pb collisions} \label{newpbpb}

\begin{figure}[h]
	\includegraphics[width=1.65in]{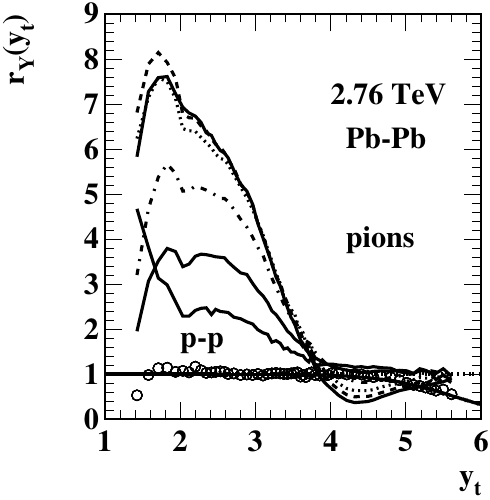}
	\includegraphics[width=1.65in]{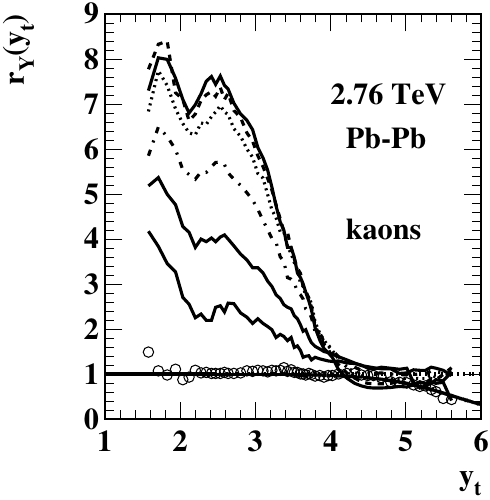}
	\put(-145,105) {\bf (a)}
	\put(-23,105) {\bf (b)}\\
	\includegraphics[width=3.3in]{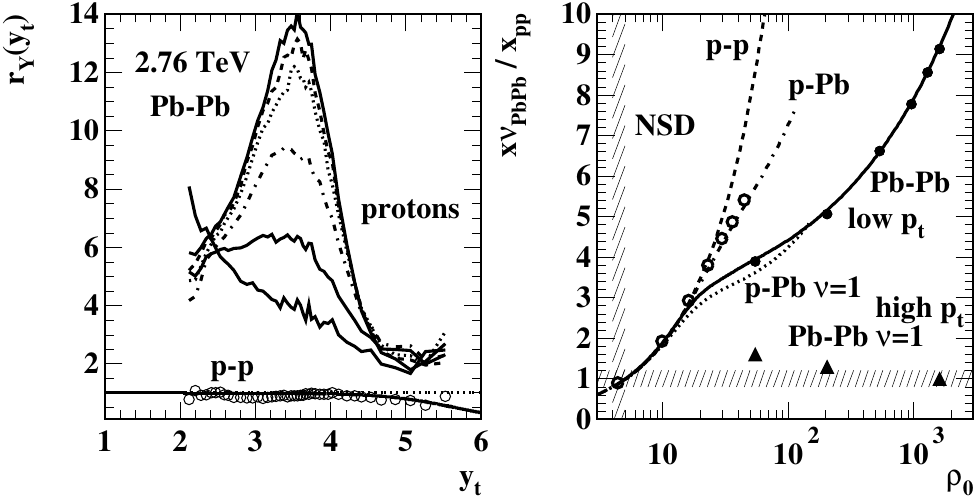}
	\put(-142,105) {\bf (c)}
	\put(-29,105) {\bf (d)}
	\caption{\label{rx}
Comparison of measure $r_{\text{Y}i}$ from 2.76 TeV \pbpb\ collisions for (a) pions, (b) charged kaons and (c) protons. The dashed lines in (a) denote the values for $\nu'$ from Table~\ref{ppbparams1}. The solid curves through \pp\ data in (a,b,c) are equivalent to curves through \pp\ data in Figs.~\ref{pions}, \ref{kaons} and \ref{protons} (d) showing systematic falloff of proton spectra. Panel (d) shows product ratio $x\nu_{\text{PbPb}}/x_{pp}$ inferred from \pbpb\ (solid points), \ppb\ (open circles) and \pp\ (dashed curve) \mmpt\ data. Solid  triangles approximate meson $r_\text{Y}(y_t)$ for \pbpb\ near \yt\ = 4.5. Note that peripheral collisions ($\bar \rho_0 < 55$) are not represented in published spectra. The pion TCM includes a resonance contribution as in Ref.~\cite{pidpart1}. Hard components are plotted here as densities on \pt\ rather than on \yt\ as in Figs.~\ref{pions}, \ref{kaons} and \ref{protons} (c).
	}  
\end{figure}

Figure~\ref{rx} shows measure $r_{\text{Y}i}$ vs \yt\ for three hadron species from 2.76 TeV \pbpb\ collisions. The $y$ axis is maintained linear to enable detailed examination of structure. These plots are therefore much more differential than conventional semilog formats such as Figs.~\ref{piddata} (right) and \ref{pions}-\ref{protons} (c,d). At lower \yt\ ratio $r_{\text{Y}i}$ varies strongly with event class corresponding to ratio $x_{\text{PbPb}}\nu / x_{pp}$ in panel (d) (solid curve), its shape consistent with the data equivalent of $\hat H_{0i\text{PbPb}}$ approximately independent of centrality. However, at higher \yt\ ratio $r_{\text{Y}i}$ is {\em approximately} consistent with 1 in (a,b).
That trend implies either that the data equivalent of $\hat H_{0i\text{PbPb}}$ is strongly suppressed relative to $\hat H_{0ipp}$ (and why should such suppression be restricted to higher-momentum fragments?) or that {\em effectively} $x_{\text{PbPb}}\nu \approx x_\text{NN}$ such that $\hat H_{0i\text{PbPb}} \approx \hat H_{0ipp}$ persists (i.e.\ no jet modification). Note that the proton $r_\text{Y}(y_t)$ trend {\em not} descending to 1 in Fig.~\ref{rx} (c) is already implicit in Fig.~\ref{protons} (b). Compare with Figs.~\ref{pions} and \ref{kaons} (b).

The $r_{\text{Y}i}$ data in Fig.~\ref{rx} then present a fundamental issue: can product $x(n_s)\nu(n_s)$ for \pbpb\ collisions be inferred independently so as to determine possible variation of the data equivalent of $\hat H_{0i\text{PbPb}}$ (jet modification) {\em assuming factorization} as noted above, or is the {\em effective} value of $x(n_s)\nu(n_s)$ subject to another degree of freedom?

The hard-component factorization in question depends on the assumption that nucleons are {\em monolithic}: A nucleon may be a participant {\em or not} ($N_{part}$), and may interact {\em a certain number of times} ($N_{bin}$) with those nucleons within an \aa\ collision partner. But that assumption conflicts with the complex structure of nucleons at high collision energies wherein a distribution of partons (i.e.\ PDF) is  relevant to participation in an A-B collision.

Paramount to such considerations is the effect of {\em exclusivity} on the nucleon PDF. Results from \pp\ and \ppb\ spectrum and \mmpt\ analysis reveal that a projectile nucleon in an established \nn\ collision cannot interact with a second nucleon ``at the same time,'' meaning within some  interval corresponding to (for instance) a nucleon diameter~\cite{tomexclude}. But that constraint may relate to relativistic time dilation and therefore to the distribution on momentum fraction $x$ of partons within nucleons. For low-$x$ partons within a parent nucleon clocks run fast and nucleons are able to respond to multiple \nn\ collisions  across a ``target'' nucleus. Ratio $x\nu_{\text{PbPb}}/x_{pp}$ then follows the solid curve and points in Fig.~\ref{rx} (d) consistent with the factorization assumption. For higher-$x$ partons time dilation may limit response to a single \nn\ interaction since a ``dead time'' corresponding to exclusivity may {\em equal or exceed a Pb diameter}. The effective $x\nu$ ratio then descends toward the hatched band. 
The trends  in Fig.~\ref{rx} are compatible with {\em nuclear transparency} as  in Sec.~\ref{transp}.

A notable feature of Fig.~\ref{rx} is the qualitative difference between meson and baryon  trends. For mesons, ratio $r_{\text{Y}i}(y_t)$ is consistent with a constant value (for each event class) below \yt\ = 3 approximating the $x\nu_{\text{PbPb}}/x_\text{NN}$ trend in panel (d). 
That \yt\ interval then suggests that ratio $\hat H_{0i\text{PbPb}} / \hat H_{0ipp}$ is consistent with 1. Above \yt\ = 3, $r_{\text{Y}i}(y_t)$ descends rapidly to 1. That could mean either that $x\nu_{\text{PbPb}}/x_\text{NN}$ remains constant (factorization) and $\hat H_{0i\text{PbPb}} / \hat H_{0ipp}$ follows such a trend (jet suppression) or that ratio $x\nu_{\text{PbPb}}/x_\text{NN}$ descends to the hatched-band limit in panel (d) due to exclusivity and time dilation.

To understand the baryon (proton in this case) $r_{\text{Y}i}(y_t)$ trends relative to mesons one should consider the role of fragmentation functions (FFs) in production of jet fragment distributions (spectrum hard components). In Ref.~\cite{ppbpid} Sec.~VI, Fig.~7 (left) summarizes 5 TeV \ppb\ spectrum hard components for pions, kaons and protons. The last is shifted to substantially higher \yt\ compared to mesons. The right panel shows measured fragmentation functions from \ee\ collisions exhibiting similar trends. Such differences are {\em inherent} in the nature of parton fragmentation to jets. The expected consequence in the present case should be that baryon jet fragments are suppressed at lower \yt, consistent with the trends in Fig.~\ref{rx} (c), above which the baryon density rises sharply to a peak near \yt\ = 3.8 ($p_t \approx 3$ GeV/c). The baryon trend then descends toward 1 consistent with meson trends. The result is a sharply-peaked proton distribution that manifests as such also in $p/\pi$ ratios and is conventionally attributed to quark coalescence or recombination~\cite{phenixptopi} and/or to radial flow~\cite{wanghydro}. One should note however that ``enhancement'' of protons near $p_t = 3$ GeV/c ($y_t \approx 3.8$) follows a similar {\em jet-related} trend (panel (d)) to that for mesons near 0.5 GeV/c ($y_t \approx 2$), but correlation between those trends is obscured by ratios such as $p/\pi$ and $R_\text{AA}$.

\subsection{5 TeV p-Pb collisions} \label{newppb}

If conventional measures such as $R_\text{AA}$ are interpreted to indicate jet modification (``jet quenching'') in \pbpb\ collisions the question then arises whether similar modification may occur in smaller collision systems, for instance \ppb\ collisions (if not \pp\ collisions!). Extensive studies of PID spectra from 5 TeV \ppb\ collisions~\cite{ppbpid,pidpart1,pidpart2} reveal small hard-component width (for mesons and baryons) and mode (for baryons) variations with event class. The variations {\em seem} quite different from those for \pbpb\ spectra as they appear in $R_\text{AA}$ results. It is therefore important to apply this same revised TCM technique to \ppb\ spectra for comparison with the \pbpb\ results above.

\begin{figure}[h]
	\includegraphics[width=1.65in]{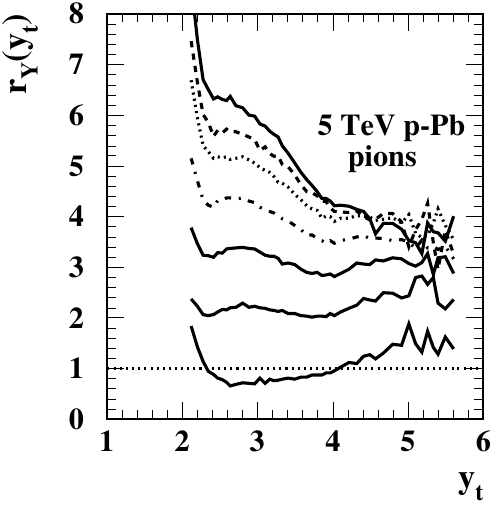}
	\includegraphics[width=1.65in]{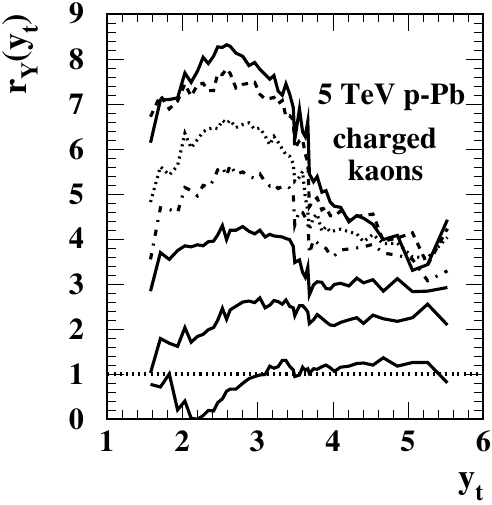}
	\put(-145,105) {\bf (a)}
	\put(-23,105) {\bf (b)}\\
	\includegraphics[width=3.3in]{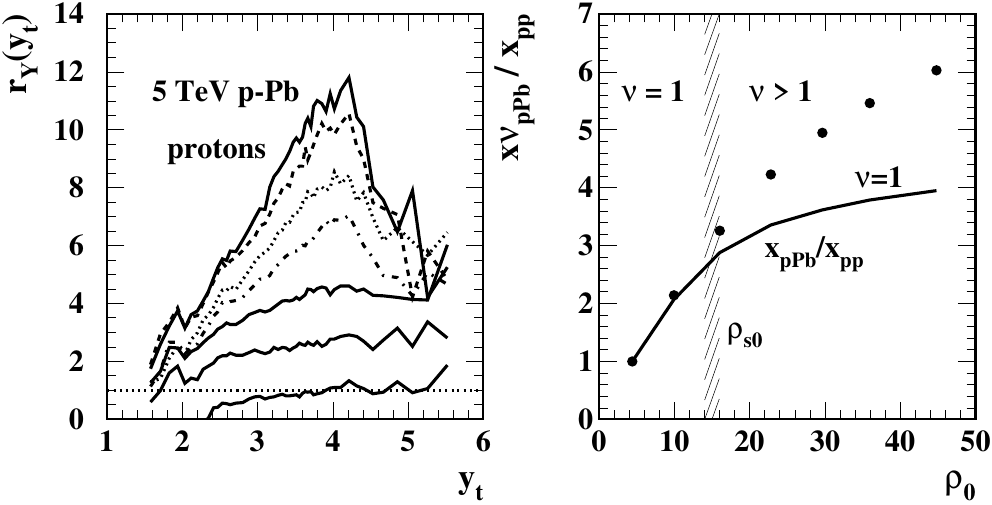}
	\put(-145,105) {\bf (c)}
	\put(-29,105) {\bf (d)}
	\caption{\label{rxx}
Hard-component ratio measure $r_{\text{Y}i}$ from 5 TeV \ppb\ collisions for (a) pions, (b) charged kaons and (c) protons. Panel (d) shows product ratio $x\nu_{\text{PbPb}}/x_{pp}$ (points) and hard/soft fraction ratio $x_{\text{PbPb}}/x_{pp}$ with $\nu = 1$ (curve). The hatched band indicates a transition point $\bar \rho_{0} \approx 15$. The pion TCM includes a resonance contribution as in Ref.~\cite{pidpart1}.
	}  
\end{figure}

Figure~\ref{rxx} shows $r_{\text{Y}i}(y_t)$ results for 5 TeV \ppb\ collisions with the same format as in Fig.~\ref{rx}. There are substantial differences from the \pbpb\ results, but they may be explained in the context of the \pbpb\ description in terms of exclusivity and time dilation. This \ppb\ description  relates to a transition near charge density $\bar \rho_{0} \approx 15$ below which exclusivity limits \ppb\ collisions to single \pn\ collisions but above which multiple \pn\ collisions are possible. The meson results at lower \yt\ correspond approximately to the $x\nu$ ratio (points) in panel (d) consistent with hard-component factorization. For three lowest event classes ($\bar \rho_0 \leq 15$) meson and baryon evolution is consistent with {\em single peripheral} \pn\ collisions arising from exclusivity ($\nu = 1$, solid curve in (d)), and $r_{\text{Y}i}(y_t)$ is approximately independent of \yt, i.e.\ the data equivalent of model $\hat H_0(y_t)$ is approximately independent of \ppb\ centrality. For four highest event classes, where \mmpt\ data indicate $\nu > 1$, lower-\yt\ meson data continue to follow the $x\nu$ ratio (points) in panel (d) while higher-\yt\ data follow the solid curve in (d) which is consistent with $\nu \approx 1$. That is, increase with $x\nu$ stalls at higher \yt\ because {\em effective} $\nu$ remains fixed at 1 and $x$ becomes slowly varying. That  result is effectively consistent with the \pbpb\ description given basic geometry differences between asymmetric and symmetric collision systems.

\vskip .2in

gaps in event spacing -- refer back to Sec. V B

\subsection{Revisited-TCM summary}  \label{revisit}

Hard/soft ratio $x \approx \alpha \rho_{s\text{NN}}$, inferred from \mmpt\ data, measures integrated jet production without regard to details of hard-component spectrum shape and thus does factorize within the full hard component. Quantity $\nu = 2N_{bin} / N_{part}$ depends on {\em incident-parton} momentum fraction $x$ via time dilation and thus  modifies the hard-component shape at higher \yt. \ppb\ collisions are dominated by hard/soft $x$ variation while $v \leq 2$ applies to any \pa\ system. \pbpb\ collisions  are dominated by $\nu$ variation while TCM $x$ may be nearly constant {\em within the limited centrality range for this analysis}. Those differences, arising from  different collision geometries,  explain the differences between Fig.~\ref{rx} and Fig.~\ref{rxx} while the responsible mechanisms -- exclusivity and parton time dilation -- play a role common to both.
It is regrettable that there is no overlap between {\em spectrum} \ppb\ and \pbpb\ \nch\ values for  these available data. Lower-\nch\ \ppb\ reveals the effects of exclusivity as in Refs.~\cite{ppbpid,tomglauber} while higher-\nch\ \pbpb\ reveals the consequences of parton time dilation as demonstrated in the present study. Taken together, a full picture of exclusivity and time dilation is {suggested} by this combination. It would be helpful to have overlap of \nch\ ranges for  the two collision systems.

\section{PID spectrum ratios}  \label{ratios}

Remarking on high-\pt\ jet quenching and PID $R_\text{AA}$ data in its Fig.~23, Ref.~\cite{alicepbpbpidspec} concludes that for $p_t > 10$ GeV/c ($y_t > 5$) three hadron species follow {\em statistically similar trends}, suggesting that while ``jet quenching'' represents in some sense parton energy loss in a dense medium the hadrochemistry (``species composition'') of the jet ``core'' is not significantly influenced by the dense medium. Concerning species/species spectrum ratios ``[t]he similarity [among species] at high \pt\ for the $R_\text{AA}$ implies that the particle ratios there are also the same in \pp\ and \pbpb\ collisions. ... We conclude that all [hadron/pion] ratios...are consistent within the systematic uncertainty....'' As noted below, those trends are indicative of manifestations of nucleon exclusivity and parton time dilation with dramatic implications for broader data interpretations.

Concerning intermediate-\pt\ phenomena and its Figs.~25 and 26 Ref.~\cite{alicepbpbpidspec} observes that ``[f]or $p_t < 10$ GeV/c, the data provide important constraints for models aimed at describing the transition from soft to hard physics.''  Within ``[t]he intermediate \pt\ regions...the proton-to-pion and the kaon-to-pion ratios are enhanced....'' The $K/\pi$ and $p/\pi$ ratios ``both show a distinct peak at $p_t \approx 3$ GeV/c...that decreases for more-peripheral collisions [i.e.\ is centrality dependent].''  ``It is an open question if additional physics processes occur in the intermediate \pt\ region 2-6 GeV/c''

The expression ``transition from soft to hard physics'' refers to a persistent expectation that the \pt\ axis may be divided into distinct intervals within which different physical mechanisms apply: ``soft'' for lower \pt, ``hard'' for higher \pt\ and an ill-defined context in between. That narrative may be strongly questioned in light of the success {\em over twenty years} of a TCM that accurately describes a large volume  of collision data based on two {\em simple and distinct} ``physics'' mechanisms that happen to overlap strongly on \pt. One may ask what is ``enhanced'' measured against? The ``distinct peak'' for proton/pion ratios is evident for reasons summarized below, but the kaon/pion ratio shows a small elevation for the most-central data that results from {\em concealing} the real kaon/pion {\em jet} peak by including spectrum soft components in the ratio. Contrast the prominent jet peaks for mesons below \yt\ = 4 (\pt\ = 3.8 GeV/c) in Fig.~\ref{rx} (a,b) with the slight rise below 4 GeV/c in the most-central panel of Fig.~25 of Ref.~\cite{alicepbpbpidspec}. As to novel physics processes, the broad success of the TCM indicates that {\em conventional nuclear physics} describes the entire \pbpb\ \pt\ acceptance.

In what follows the TCM, including new insights gained from the present study, is applied to several spectrum ratios from Ref.~\cite{alicepbpbpidspec}. Discussion is based on TCM versions of the respective ratios as described below.
A TCM version of ratio $R_\text{AA}$ as it appears in Fig.~23 of Ref.~\cite{alicepbpbpidspec} is given by Eq.~(\ref{raann}) (first line). The second line gives a simplified form in case \pbpb\ collisions are approximately linear superpositions of \nn\ collisions.
Based on TCM definitions in Sec.~\ref{spectrumtcm},  spectrum ratios in Figs.~25 ($K/\pi$) and 26 ($p/\pi$) (upper panels) of Ref.~\cite{alicepbpbpidspec} are described at higher \yt\ ($y_{t\pi} = \ln[(p_t+m_{t\pi})/m_\pi]$) by
\bea \label{ratioeq}
\frac{\bar \rho_{0i}(y_t)}{\bar \rho_{0j}(y_t)} &\rightarrow & \frac{\tilde z_i}{\tilde z_j} \cdot \frac{ x \nu \hat H_{0i}(y_t)}{ x \nu \hat H_{0j}(y_t)}  \cdot \frac{z_{si}(n_s)}{z_{sj}(n_s)} 
~~
\\ \nonumber
&\approx& \frac{\tilde z_i}{\tilde z_j} \cdot \frac{\hat H_{0i}(y_t)}{\hat H_{0j}(y_t)} \cdot \frac{z_{0i}(1+\tilde z_j x\nu)}{z_{0j}(1+\tilde z_i x\nu)},
\eea
where $i$ is kaons or protons and $j$ is pions. Those expressions are used to interpret  Figs.~23, 25 and 26 of Ref.~\cite{alicepbpbpidspec}.

\subsection{\bf $\bf R_\text{\bf AA}(y_t)$ ratios}

With regard to its Fig.~23 Ref.~\cite{alicepbpbpidspec} states that ``...for all event classes any particle species dependence...for $p_t > 10$ GeV/c is small, compared with the large suppression ($R_\text{AA} \ll 1$). This suggests that jet quenching does not produce signatures  that affect the particle species composition for leading particles.'' After some discussion of fragmentation functions (FFs) it is concluded that ``[t]he results in Fig.~23 therefore indicate  that for jets with final \pt\ of order 25 to 50 GeV/c, jet quenching does not produce large particle-species-dependent effects in the hard core of the jet where leading particle production mainly occurs.''
In each panel (i.e.\ \pbpb\ centrality) of that figure, \pbpb\ and \pp\ spectra are simply compared in ratio for three hadron species. The definition of $N_{bin}$ or interpretation of $R_\text{AA}$ {\em per se} is then not relevant.

It is  instructive to compare Fig.~23 of Ref.~\cite{alicepbpbpidspec} with some details of Fig.~\ref{rx} from the present study. For mesons in Fig.~\ref{rx} (a,b)  ratio $r_\text{Y}(y_t)$ is approximately 1 above \yt\ = 4 ($p_t \approx 3.8$ GeV/c). That translates to $R_\text{AA}$ equivalence for those meson species within the same \pt\ interval corresponding to Fig.~23 of Ref.~\cite{alicepbpbpidspec}. For protons (baryons) in panel (c) ratio $r_\text{Y}(y_t)$ is approximately 1 above \yt\ = 4.75 ($p_t \approx 8$ GeV/c) which again matches the trends in Fig.~23 of Ref.~\cite{alicepbpbpidspec} for protons vs pions. Thus, for hadrons relative to pions, kaons are consistent with pions above 4 GeV/c and protons are consistent with pions above 8 GeV/c agreeing with Fig.~23 of Ref.~\cite{alicepbpbpidspec} as noted. Figure~\ref{rx} then {\em anticipates} the information conveyed in Fig.~23 of Ref.~\cite{alicepbpbpidspec} but in a {\em more interpretable form}. As discussed in Sec.~\ref{extend} the structure of Fig.~\ref{rx} that corresponds to the trends in Fig.~23 of Ref.~\cite{alicepbpbpidspec} noted above may be attributed to a combination of exclusivity and time dilation as experienced by participant {\em partons}.

Evolution of structure at lower \pt\ {\em relative to} higher \pt\ in Fig.~23 is explained as follows. At lower \pt\ in Fig.~\ref{rx} jet contributions strongly increase per quantity $x\nu_{\text{PbPb}}/x_{pp}$ (solid curve) in panel (d). Although the mode of the proton peak near \yt\ = 3.8 is much higher than that for meson peaks near \yt\ = 2.5 the amplitude trends are {\em very similar}. In Fig.~23 of Ref.~\cite{alicepbpbpidspec} the meson peaks appear much reduced compared to the proton peaks, but that difference arises from a fundamental problem with spectrum ratios that {\em include soft components}. Those soft components dominate  ratios at lower \pt\ (below $\hat H_0(y_t) / \hat S_0(y_t) \approx 1$) and {\em overwhelm the jet contribution}, much more so for meson peaks with modes at lower \yt. In Fig.~\ref{rx}, where only hard components appear in ratio, the large meson jet contributions at lower \yt\ are clearly evident.

Another aspect of Fig.~23 is of interest. The $R_\text{AA}$ trends appear to increase substantially and monotonically above \pt\ = 8 GeV/c. But note that the \pp\ trends in Fig.~\ref{rx} for all three hadron species {\em decrease} substantially (tens of percent, lowest solid curves) relative to a $\hat H_0(y_t)$ reference including a simple exponential tail on \yt\ (power-law tail on \pt) that describes {\em most other} \pp\ spectrum data. Whereas quantity $r_{\text{Y}}(y_t)$ includes $\hat H_0(y_t)$ as a reference in its denominator, $R_\text{AA}$ includes \pp\ {\em data} spectra as a reference in its denominator leading to the monotonic increase noted above. Given experience with other \pp\ spectrum data it is possible that the \pp\ data in this case are biased, producing misleading $R_\text{AA}$ variation.

\subsection{$\bf K/\pi$ ratios}

With regard to its Fig.~25 Ref.~\cite{alicepbpbpidspec} states that, since in that figure particle ratios are constant on \pt\ above 6 GeV/c, ``we use the integrated particle ratios for $p_t > 10$ GeV/c [plotted in its Fig.~27] to elucidate the precision with which the suppression of pions, kaons and protons is similar....'' ``We conclude that all kaon-to-pion ratios [vs $N_{part}$, see below for proton-to-pion] are consistent within the systematic uncertainty....''

Just as for its Fig.~23 we can compare Fig.~25 of Ref.~\cite{alicepbpbpidspec} with some details of Fig.~\ref{rx} from the present study. In panels (a,b) of the latter figure pions and kaons from \pbpb\ collisions are approximately equivalent to those from \pp\ collisions ($r_\text{Y}(y_t) \approx 1$) for \yt\ $> 4$ ($p_t > 3.8$ GeV/c). That implies the kaon/pion ratio should be independent of \yt\ {\em and} \pbpb\ centrality over the same interval, agreeing with Fig.~25 of Ref.~\cite{alicepbpbpidspec}. But via the TCM for identified-hadron spectra a {\em specific value} of the ratio can be predicted based on parameters derived from previous analysis of \pp\ and \ppb\ spectra and Eq.~(\ref{ratioeq}).

Equation~(\ref{ratioeq}) includes hard/soft abundance ratios $\tilde z_i$ and total fractional abundances $z_{0i}$ (that relate to statistical models). Based on Fig.~\ref{tildezparams} the $\tilde z_i$ values are 0.6 and 2.6 for pions and kaons respectively. Fractional abundances $z_{0i}$ are 0.8 and 0.125 respectively. As noted, from Fig.~\ref{rx} (a,b) $\nu \approx 1$ for $y_t > 4$ ($p_t >3.8$ GeV/c) which corresponds to the interval for Fig.~25 (upper panels) of Ref.~\cite{alicepbpbpidspec} where the spectrum ratio is approximately uniform. The ratio $\hat H_{0i}(y_t) / \hat H_{0j}(y_t)$ should be approximately 1 over the same \yt\ interval {\em if} the tail parameter $q \approx 4$ is the same for pions and kaons as in Table \ref{pidparams} for \ppb\ spectra. Over the centrality range of \pbpb\ spectrum data for this study $x \approx 0.16$ is based on a TCM analysis of \mmpt\ data as in Sec.~\ref{altgeom}.
Those numbers in Eq.~(\ref{ratioeq}) lead to constant value $\bar \rho_{0i}(y_t)/\bar \rho_{0j}(y_t) \approx 0.50\pm 0.05$ for $p_t > 3.8$ GeV/c that agrees well with the $K/\pi$ spectrum ratios reported in Fig.~25 of Ref.~\cite{alicepbpbpidspec}. 

Regarding intermediate-\pt\ results Ref.~\cite{alicepbpbpidspec} refers to hadron/pion ratio {\em enhancements} (presumably referring to increases relative to some reference level) but the reference value for such observations is not made clear. Such enhancements are conventionally attributed to quark recombination~\cite{phenixptopi} and/or to hydrodynamical flow~\cite{wanghydro}.

Once again  Fig.~\ref{rx} from the present study sheds light on this matter, focusing now on the pion (a) and kaon (b) trends below \yt\ = 4 ($p_t \approx 3.8$ GeV/c) where there are large {\em jet-related} contributions in those panels. As noted above, for spectrum ratios in which {\em soft components are included jet contributions are overwhelmed} at lower \pt. That is particularly evident in Fig.~\ref{piddata} (d,e) with the steep descent below jet-related peaks near $p_t = 2$ GeV/c. For the $K/\pi$ ratio there is a further sharp reduction in the jet peak because, as is evident in Fig.~\ref{rx} (a,b) and in Figs.~\ref{pions} and \ref{kaons} (c), those hard-component peak shapes are quite similar and thus {\em effectively cancel in the ratio}. What one then observes in Fig.~25 (upper panel) of Ref.~\cite{alicepbpbpidspec} below 4 GeV/c is a {\em vestigial remainder} of the lower-\yt\ strong jet contribution in Fig.~\ref{rx} (a,b) after a {\em double cancellation}. An electrical engineer might refer to common-mode noise rejection by subtraction or ratio of two channels with correlated noise sources. In the present instance the ratio of two ``channels'' results in common-mode {\em signal} rejection.

\subsection{$\bf p/\pi$ ratios}

As noted by Ref.~\cite{alicepbpbpidspec}, $p/\pi$ spectrum ratios have been the subject of intense speculation since early in the RHIC program as possibly signaling novel mechanisms relating to a QGP, including constituent-quark recombination and possible coupling of flow and jet suppression. However, again in the context of Fig.~\ref{rx} the seemingly contrasting structure of Figs.~25 and 26 is easily explained.

Differences are accounted for by the different properties of meson and baryon hard components. The latter are shifted to higher \yt\ and substantially narrower, with more rapid falloff than the former. Compare peak shapes in Fig.~\ref{pions} (c) and \ref{protons} (c). In Table~\ref{pidparams} compare $\bar y_t$ (modes greater for baryons) and $\sigma_{y_t}$ and $q$ (widths and exponential tails descending more rapidly for baryons). Those differences are consistent with {\em measured} fragmentation functions for identified hadrons from unidentified light quarks and gluons as noted in Sec. VI C of Ref.~\cite{ppbpid}.

At higher \pt\ in Fig.~26 (upper panels) there is no significant centrality dependence (\pbpb\ $\approx$ \pp), consistent with trends for $r_{\text{Y}_i}(y_t) \approx 1$  above \yt\ = 4.75 or $p_t = 8$ GeV/c for $p/\pi$ in Fig.~\ref{rx}. However, unlike the $K/\pi$ (meson/meson) ratio the $p/\pi$ (baryon/meson) ratio is strongly \pt\ dependent because of different widths and tails of pion and proton hard components as noted above.

At lower \yt\ the $p/\pi$ ratio shows strong centrality dependence, in sharp contrast to near independence of the $K/\pi$ ratio. The difference is mainly due to (a) mass dependence of the hard/soft fractional-abundance ratio $\tilde z_i$ and (b) different meson and baryon hard-component mode locations. 
The relation can be made clear via the thin reference curves in Fig.~\ref{piddata} (d vs f) vs $R_\text{AA}$ data. The reference curves assume $x_\text{PbPb} \approx x_{pp}$ (\nn\ linear superposition) whereas for data $x_\text{PbPb} \propto \bar \rho_{s\text{PbPb}}$ increases sharply with $\bar \rho_0$ as in Fig.~\ref{story} (c). What determines the reference rate of rise at lower \yt, is $\tilde z_i$ -- 0.6 for pions vs 5.6 for protons -- thus accounting for the very different corner positions in Fig.~\ref{piddata} (right) at 5 GeV/c for pions vs 2.5 GeV/c for protons. 
In contrast, for actual {\em data} hard components in Fig.~\ref{rx} mesons peak below \yt\ = 2.7 ($p_t \approx 1$ GeV/c) whereas protons peak near \yt\ = 3.8 ($p_t \approx 3$ GeV/c). Thus, for both $R_\text{AA}$ and $p/\pi$ spectrum ratios protons exhibit a dominant peak whereas mesons {\em appear} strongly suppressed. The difference results simply from {\em measured} hadron mass dependences of $\tilde z_i$ and FFs.

\section{Systematic uncertainties} \label{sys}

Analysis of \pbpb\ PID spectra in the present analysis has been carried out in the context of previous analyses of \pp\ \pt\ spectra~\cite{ppprd,ppquad,alicetomspec} and PID \ppb\ spectra~\cite{tommpt,tomglauber,tomexclude,ppbpid}. In particular, the 5 TeV \ppb\ PID spectrum TCM reported in Ref.~\cite{ppbpid} provides a basis for the present analysis. Uncertainties in primary data and basic TCM analysis are presented there. However, uncertainty estimation should extend beyond those for primary particle data. Uncertainties arising from plotting formats, applied data models and physical interpretations of data structure and models are also relevant. This section emphasizes two examples: (a) some consequences of Glauber estimation of \pbpb\ collision geometry and (b) factorization of spectrum hard components.

\subsection{Pb-Pb collision geometry and low-$\bf y_t$ data} \label{pbpbgeom}

To extract spectrum hard components accurately within the TCM context soft rescaling of published \pt\ spectra to obtain quantity $\text{X}_i(y_t)$ requires accurate estimation of soft-component density $\bar \rho_{si}(y_t)$ so as to match model function $\hat S_{0i}$ at low \yt, with
\bea \label{rhosii}
\bar \rho_{si} = \frac{z_{0i}\bar \rho_{0}}{1+\tilde z_i x \nu}.
\eea
That problem relates to the structure of Eqs.~(\ref{pidspectcm}), (\ref{rhosi}) and (\ref{pidhard}) above. Detailed study of 5 TeV \ppb\ spectra in Refs.~\cite{ppbpid,pidpart1,pidpart2} established that fractional abundances $z_{0i}$ are consistent with statistical-model predictions and $\tilde z_i$ is simply proportional to hadron mass. In Refs.~\cite{ppbpid,tomglauber} it was established that Glauber-model estimates of $N_{part}$ and $N_{bin}$ greatly overestimate those quantities for \ppb\ collisions. An alternative is provided by inverting \mmpt\ data to obtain product $x\nu$ as described in Sec.~\ref{altgeom} above. It is of interest to consider  the consequences of using conventional \pbpb\ Glauber estimates for the present study.

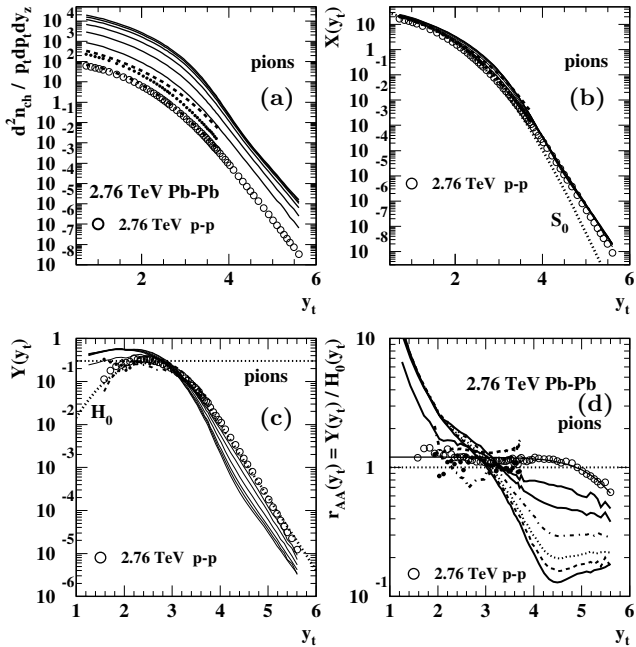
\begin{figure}[h]
	\includegraphics[width=3.3in]{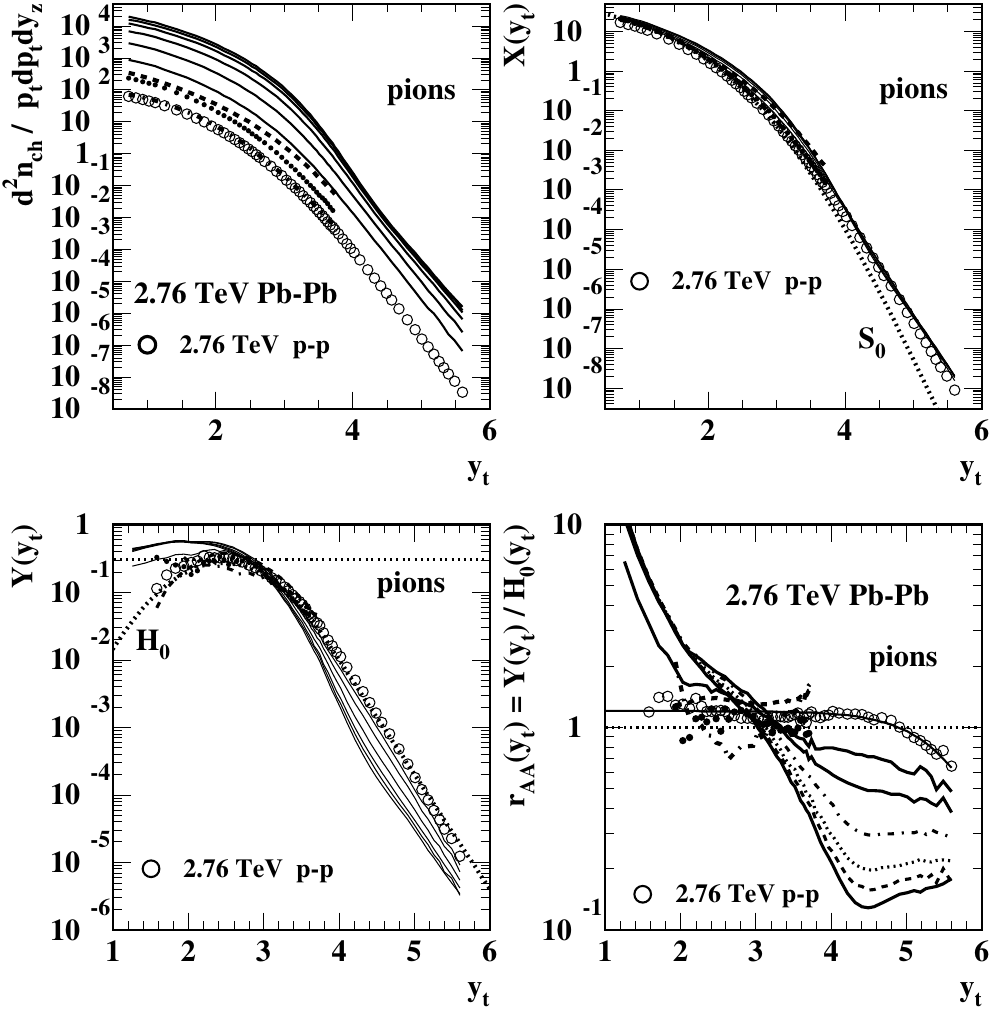}
\put(-142,205) {\bf (a)}
\put(-25,205) {\bf (b)}
\put(-142,84) {\bf (c)}
\put(-22,89) {\bf (d)}
	\caption{\label{pionsbad}
		(a) Pion spectra from six event classes of 2.76 TeV \pbpb\ collisions as published (solid curves)~\cite{alicepbpbpidspec}  plotted on pion transverse rapidity $y_{t\pi}$. Spectra from midcentral (dashed) and peripheral (dash-dotted) 5 TeV \ppb\ collisions are included as a reference. Also included are peripheral 80-90\% central \pbpb\ data (solid points) and \pp\ data (open circles).
		(b) Data from panel (a) plotted in format $\text{X}(y_t)$ defined by the third line of Eq.~(\ref{pidspectcm}). 
		(c)  Data from panel (b) plotted in format $\text{Y}(y_t)$ defined by Eq.~(\ref{pidhard}).
		(d)   Data from panel (c) plotted as hard-component ratios $r_\text{AA}(y_t)$ defined by Eq.~(\ref{raax}).
	}  
\end{figure}

\begin{figure}[h]
	\includegraphics[width=3.3in]{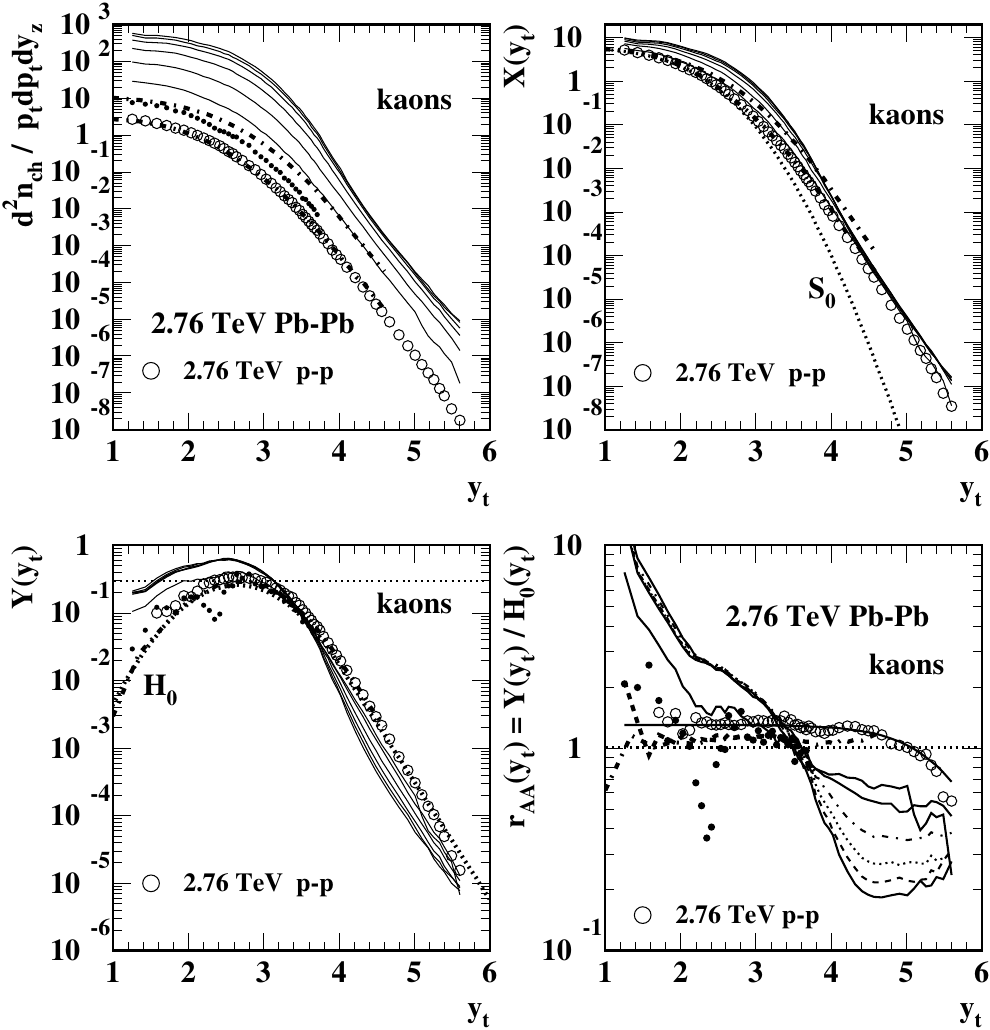}
\put(-142,205) {\bf (a)}
\put(-25,205) {\bf (b)}
\put(-142,85) {\bf (c)}
\put(-22,75) {\bf (d)}
	\caption{\label{kaonsbad}
Kaon spectra from 2.76 TeV \pbpb\ and \pp\ collisions ($K^\pm$) and 5 TeV \ppb\ collisions ($K^\pm,~K^0_\text{S}$) plotted as in Fig.~\ref{pions}. In panels (b) and (c) only $K^0_\text{S}$ \ppb\ data are plotted.
	}  
\end{figure}
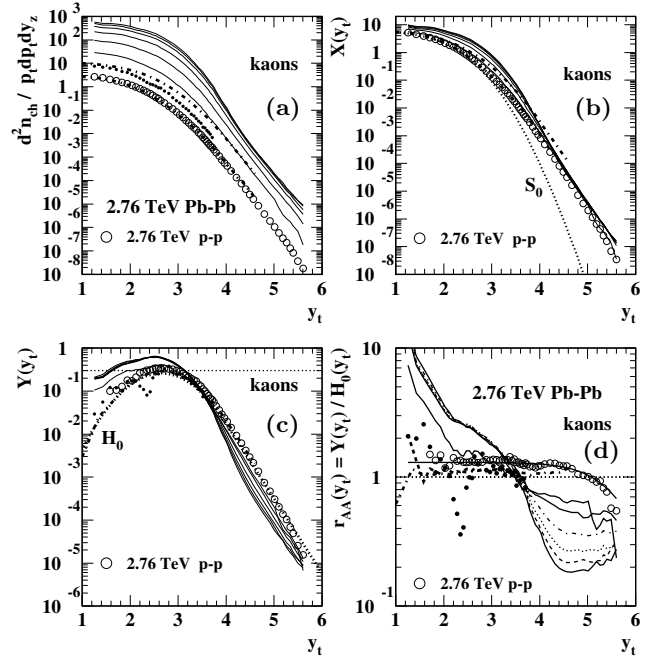

Figures~\ref{pionsbad} and \ref{kaonsbad} demonstrate the result. Contrast those figures with Figs.~\ref{pions} and \ref{kaons} that  employ \mmpt\ estimates for $x\nu$. While the jet contribution for pions is relatively low ($\tilde z_i$ is proportional to mass) the contribution for kaons is sufficient to reveal in panel (b) that there is up to a factor 2 discrepancy between rescaled data and $\hat S_0$ at \yt\ = 1.3 ($p_t \approx 0.24$ GeV/c). The origin of the systematic difference is easy to determine: Fig.~\ref{inverse} (right) shows $x\nu$ from the Glauber estimate a factor two higher for the most-central event class than that obtained from \mmpt\ data. Equation~(\ref{rhosii}) indicates that $x\nu$ too large drives down the estimate for $\bar \rho_{si}$ which in turn drives up rescaled $\text{X}_i(y_t)$.

Such systematic errors matter little for the presentation format of Fig.~\ref{piddata} since $R_\text{AA}(p_t)$ (right panels) strongly suppresses low-\pt\ results (i.e.\ most jets), making them visually inaccessible. However, in Figs.~\ref{pionsbad} and \ref{kaonsbad} the low-\yt\ \pbpb\ results in panels (c) and (d) {\em strongly deviate} (factor 10) from  the \pp\ and \ppb\ results that agree with model $\hat H_0(y_t)$. If one accepts  the Glauber results one must account for the very strong \pbpb\ ``enhancement'' at lowest \pt, a very large excess of hadron production.

\subsection{Factorization of spectrum hard components}

Section~\ref{extend} introduces an alternative TCM approach to spectrum hard components that arises from concerns over assumptions about hard-component factorization that are also implicit for $R_\text{AA}$ (e.g.\ assuming that $N_{bin}$ can be isolated as a separate factor). This situation illustrates a {\em combination} of uncertainties including primary data quality, model formulation, plotting format and interpretation that relate to spectrum hard components isolated in the form $\text{Y}'(y_t)$ as in Eq.~(\ref{yprime}). 

$R_\text{AA}(y_t)$ is motivated by and plotted to feature ``jet suppression'' at higher \yt\ where jets are expected to dominate. $r_\text{AA}(y_t)$ is formulated to make available {\em all} jet contributions, especially at lower \yt, but still reflects a context of jet suppression however that might appear. The revised formulation in Sec.~\ref{extend} is motivated to retain the low-\yt\ accessibility feature of $r_\text{AA}(y_t)$ but abandons the factorization assumption that has become doubtful.
Intact hard components presented in a revised plot format suggest new interpretations for their properties and clarifies the importance of exclusivity and time dilation. 

Introduction of a revised TCM system as in Sec.~\ref{extend} leads to dramatically different jet-related data interpretation: $N_{bin}$ and $\nu$ relate to partonic, not hadronic, degrees of freedom. Parton speed (i.e.\ nucleon momentum fraction $x$) then determines time dilation and thus the effect of empirically-observed exclusivity that determines {\em effective} $\nu$ varying (with jet fragment \pt) from projectile-nucleon $\nu$ (inferred from \mmpt\ data) at lower $x$ to 1 at higher $x$. Revised interpretation of \pbpb\ data is buttressed by correspondence with early fixed-target \pa\ experiments demonstrating {\em nuclear transparency} via  distributions on pseudorapidity $\eta$ and their variation with target A~\cite{buszanew}.

\section{Discussion} \label{disc}

 This section considers three issues that emerge from a TCM analysis of 2.76 TeV \pp\ and \pbpb\ PID spectrum data in the context of previous TCM analysis:
nuclear transparency at the LHC and FNAL,
complementarity of \ppb\ vs \pbpb\ systems,
and relation of a conjectured dense flowing medium to jet production.

\subsection{Nuclear transparency at the LHC and FNAL} \label{transp}

The trends observed in Sec.~\ref{newpbpb} appear compatible with the discovery of ``nuclear transparency'' at Fermilab in the 1970s. Compare Figs.~8 and 9 of Ref.~\cite{buszanew} with meson panels (a) and (b) of Fig.~\ref{rx}. The former show distributions on pseudorapidity $\eta \sim y_z$ of particle densities from \pa\ collisions for several targets A including H$_2$. Particle densities in the target hemisphere (low $y_z$) increase with target A (measured by $\bar \nu \sim N_{bin}$) while densities in the projectile-proton hemisphere (high $y_z$) are independent of A (transparency, i.e.\ consistent with target  H$_2$). The latter show distributions on transverse rapidity \yt\ of jet-related particle (fragment) densities where, for lower \yt, particle densities (for mesons at lease) follow a $x\nu \rightarrow \nu$ trend while at higher \yt\ particle densities are consistent with single \pp\ (\nn) collisions with $\nu \approx 1$. 

The two systems can be related by interpreting the $\eta$ distributions in the \pa\ figures of Ref.~\cite{buszanew} to reflect the constraints of time dilation and exclusivity on parton interactions along the beam axis. Among partons that do interact (in \pa\ {\em or} \aa) some may scatter at large angles to produce jet-related particle distributions on \yt. In essence, a parton distribution on  projectile $x\rightarrow y_z$ is mapped onto a fragment distribution on $p_t \rightarrow y_t$ with properties reflecting  similar exclusivity and time-dilation constraints in each context. What appears to be demonstrated in Sec.~\ref{newpbpb} is  persistence of nuclear transparency at LHC energies and even in \pbpb\ collisions.

\subsection{Complementarity: p-Pb vs Pb-Pb systems} \label{complement}

It may be useful to describe the complementary interplay between \ppb\ and \pbpb\ analyses in developing a revised TCM analysis. Possible proton inefficiency in LHC  \pbpb\ spectra was already noted in a study of \auau\ and \pbpb\ {\em quadrupole spectra} (derived from $v_2$ data) in 2016~\cite{njquad}. Work on 5 TeV \ppb\ spectra in 2018 included a study of a \ppb\ Glauber model \cite{tomglauber},  discovery of exclusivity~\cite{tomexclude} and the first PID TCM for \ppb\ collisions~\cite{ppbpid}.
\ppb\ analysis provided insight as to geometry determination via \mmpt\ data and, because of \ppb\ simplicity, provided an exact solution that could be tested against spectrum data. That initial PID spectrum description included the concept of species fractional abundances $z_{xi}(n_s)$.

Return to work on \pbpb\ spectrum paper commenced in late 2019 but Glauber geometry issues and systematic errors in $dE/dx$ PID analysis led to postponement. A first correction for proton $dE/dx$ bias was established.

A followup TCM analysis of PID \ppb\ spectra in 2021 ~\cite{pidpart1,pidpart2} provided complete details on PID spectrum evolution and the discovery of mass-dependent conserved transport of hadron species from soft to hard component in 2022~\cite{transport} permitting precise correction of proton spectra in \ppb.
However, examination of \ppb\ spectrum hard components for evidence of jet modification was inconclusive: small width and centroid variations could be accurately modeled but provided little physical insight.

Application of the revised \ppb\ PID TCM to \pbpb\ data in late 2024 revealed two important features: (a) the lower-\yt\ parts of spectrum hard components (jet fragments) show no deviation from TCM predictions (no suppression) whereas higher-\yt\ parts show {\em apparent} strong suppression (as in $R_\text{AA}$ results) and (b) spectra rescaled per Eq.~(\ref{pidspectcm}) (based on \mmpt\ from \pbpb\ analysis) appear {\em precisely congruent to \pp\ spectra at higher \yt} unlikely to result from suppression by a dense medium. That combination led to questioning assumed hard-component {\em factorization} that is the basis for both conventional measure $R_\text{AA}$ and previous TCM measure $r_\text{AA}$. Removing the factorization assumption led to plotting hard components in ratio to the simplest possible \nn\ reference using a linear vertical axis as in Fig.~\ref{rx} that maximizes precision. Whereas the \pbpb\ result shows a dramatic transition from lower \yt\ to higher \yt\ the \ppb\ results in Fig.~\ref{rxx} show a nuanced variation that has led to deeper insight. Lack of overlap between \ppb\ data (open circles) and \pbpb\ data (solid dots) in Fig.~\ref{rx} (d) illustrates the impact at present of insufficient continuity on centrality or \nch\ of the two collision systems. Nevertheless, the value of comparing the two systems during development of the analysis system (complementarity) is illustrated.

\subsection{Dense flowing medium vs jet production} \label{where}

Interpretation of \pbpb\ PID \pt\ spectra presented in Ref.~\cite{alicepbpbpidspec} is summarized in Sec.~\ref{ratios}. Two \pt\ intervals are  assumed to reflect distinct physical mechanisms: For $p_t < 3$ GeV/c spectrum features are assumed to reflect ``bulk production'' (including flow manifestations), and for $p_t > 10$ GeV/c jet formation and in-medium jet modification (``jet quenching'') should be the relevant issue for produced hadrons. If that description were valid jet fragments would comprise a tiny fraction of hadron production and almost all hadrons (i.e.\ most of the {\em nominal} nonjet fraction) would emerge from a flowing dense medium (QGP). A similar interpretation of \ppb\ PID \pt\ spectra is presented in Ref.~\cite{aliceppbpid}. Between separate low-\pt\ and high-\pt\ intervals lies an intermediate region where the hadron production mechanism is ill-defined. Such descriptions are based on the assumption that distinct production mechanisms relate to separated \pt\ intervals.

In contrast to spectrum models based on such {\em a priori} assumptions the TCM is based on {\em observed} scaling of distinct soft and hard hadron fractions that happen to overlap strongly on \pt. Derivation of the \pp\ TCM from \nch\ systematics of \pt\ spectra is first described in Ref.~\cite{ppprd}. Accurate isolation of the two spectrum components then leads to direct comparison between the spectrum hard component and measured jet properties~\cite{fragevo} and to development of an accurate description of the \pp\ collision-energy evolution of jet energy spectra over  three orders of magnitude~\cite{jetspec2}. In essence, measured jet spectra and measured jet FFs predict the shape and absolute magnitude of the spectrum hard component for any \pp\ collision energy. In particular, predicted \pp\ jet fragment distributions all have modes near 1 GeV/c, and most jet fragments appear {\em below} 3 GeV/c. The spectrum TCM soft component depends only on hadron mass and collision energy and does {\em not} vary with A-B centrality

With the spectrum hard-component form predicted quantitatively from measured jet properties~\cite{fragevo} the spectrum interpretations in Refs.~\cite{aliceppbpid} and \cite{alicepbpbpidspec} may be reconsidered. Describing the lower-\pt\ interval  ($< 3$ GeV/c) is the statement  ``hardening of the spectra'' with increasing centrality ``...is mass dependent and is characteristic of hydrodynamic flow....'' That qualitative observation does not establish  that flow plays any role in nuclear collisions. In fact, the {\em predicted} trend for the MB jet contribution to spectra has exactly those characteristics: (a) With increasing A-B centrality the jet contribution increases much faster than the soft component, thus making the spectrum ``harder,'' and (b) as demonstrated for \ppb\ collisions in Ref.~\cite{ppbpid} the hard/soft (jet/nonjet) abundance ratio $\tilde z_i$ is simply proportional to hadron mass for pions, kaons and protons as presented above in Fig.~\ref{tildezparams}. 

Describing a higher-\pt\ interval ($> 10$ GeV/c) is the statement ``the spectra follow a power-law shape as expected from perturbative QCD (pQCD) calculations.'' But pQCD theory is not required to predict the high-\pt\ trend derived quantitatively from the convolution of a measured jet spectrum~\cite{jetspec2} with measured FFs~\cite{eeprd}. A power-law trend on \pt\ only approximates, over a limited \pt\ interval, the actual underlying jet energy spectrum~\cite{jetspec2,mbdijets}. The same convolution predicts jet fragment distributions down to 0.5 GeV/c {\em or below} and confirms that most jet fragments appear near 1 GeV/c~\cite{hardspec,fragevo}.

Regarding an intermediate-\pt\ region (e.g.\ 2-10 GeV/c) is the statement ``...it is an open question if additional physics processes [e.g.\ the  ``baryon/meson puzzle'' \cite{rudy,duke,tamu}] occur in the intermediate $p_T$ region...''~\cite{alicepbpbpidspec}. Reference~\cite{alicespec2} states that ``Below the [proton/pion] peak [on \pt], $p_t < 3$ GeV/c, both ratios [$p/\pi$, $K/\pi$] are in good agreement with hydrodynamical calculations, suggesting that the peak itself is dominantly the result of radial flow....'' But Sec.~\ref{newpbpb} Fig.~\ref{rx} demonstrates that the proton peak (c) and related meson structures (a,b) are strongly correlated in amplitude with each other and with hard/soft ratio $x\nu$ inferred from \mmpt\ data (d). Spectrum {\em variation} in that \pt\ interval is thus dominated by jet production.

\section{Summary}\label{summ}

This article describes differential analysis of  identified-hadron (PID) \pt\ spectra from six centrality or \nch\ classes of 2.76 TeV \pbpb\ collisions and \pp\ collisions.  The object of study is evidence for jet suppression in small and large collision systems as indicating quark-gluon plasma (QGP) formation there. 
 PID spectra are described by a two-component (soft + hard) model (TCM)  of hadron production in high-energy nuclear collisions. The soft component is associated with longitudinal projectile-nucleon dissociation, and the hard component is associated with large-angle scattering of projectile partons to form jets. Previous 5 TeV \ppb\ analysis establishes TCM model functions and PID parameters for five hadron species that are adopted with minor changes for \pbpb\ pion, charged-kaon and proton spectra. Collision geometry (centrality parameters $N_{part}$ and $N_{bin}$) is derived from ensemble-mean \mmpt\ data, not a Glauber Monte Carlo.

Standard practice for TCM analysis in the past has assumed factorization of each of soft and hard components into a particle-density factor and a unit-normal model function. Implementation of that procedure in this study leads to observation of apparent strong suppression (jet modification) of \pbpb\ hard components {\em above} $p_t \approx 1.5$ GeV/c contrasting with results for 5 TeV \ppb\ spectra where evidence for hard-component suppression is inconclusive. The \pbpb\ result seems to correspond with conventional ratio $R_\text{AA}$ at higher \pt\ but strongly disagrees at lower \pt, in part because of different measure definitions.

The present study newly reports {\em precise congruence} of \pbpb\ spectra with \pp\ spectra (especially for mesons) above $p_t \approx 4$ GeV/c after \pbpb\ spectra have been rescaled by soft density $\bar \rho_{si}$ for  hadron species $i$. That observation, plus the apparent lack of suppression of hard components {\em below} 1.5 GeV/c, leads to reconsideration of the hard-component factorization assumption.

In an alternative procedure, intact \pbpb\ hard components with no factorization are compared in ratio with accurate models of \pp\ hard components, the ratio denoted by quantity $r_{\text{Y}}(y_t)$. Resulting ratio trends (for mesons) consist of general increase with centrality below $p_t \approx 4$ GeV/c, consistent with \mmpt\ data, but decreasing to fixed value 1 above 4 GeV/c. Proton results are qualitatively similar.
Given the updated data presentation format and results from \ppb\ studies these new results suggest a remarkably different physical interpretation. 

Previous analysis of 5 TeV \ppb\ spectrum data led to the conclusion that {\em exclusivity} is an essential feature of \nn\ collisions within A-B collisions: A participant nucleon can interact with only one other participant {\em at a time} as part of a quantum transition. A Glauber Monte Carlo simulation including that constraint comes much closer to describing \ppb\ collision geometry. In the present \pbpb\ study it becomes clear that nucleons cannot be treated as monolithic collision partners and that ``at a time'' relates to partons within participant nucleons that span a broad distribution of speeds (momentum fractions $x$) and therefore experience a comparable range of {\em relativistic time dilations}. Thus, ``at a time'' applied to low-$x$ partons (the majority, contributing to lower-\pt\ hadrons) may correspond to multiple \nn\ collisions per participant nucleon whereas the same measure applied to high-$x$ partons (contributing to higher-\pt\ hadrons) may correspond to a distance larger than a Pb-nucleus diameter, thus permitting only a single \nn\ collision for a corresponding participant nucleon within \pbpb.

Implications arising from these results are as follows: 
(a) Even for central \pbpb\ collisions the number of participant nucleons is about 1/3 that estimated by a classical Glauber Monte Carlo, in part because of exclusivity for \nn\ collisions as quantum transitions.
(b) Spectrum hard components do not factorize as is conventionally assumed (e.g.\ $R_\text{AA}$), in part because $N_{bin}$ or $\nu$ actually relates to individual projectile partons and depends then on time dilation corresponding to momentum fraction $x$. The {\em effective} value of $N_{bin}$ relating to produced hadrons is thus strongly \pt\ dependent.
(c) Hadron production corresponding to high-$x$ partons is restricted to single \nn\ collisions for corresponding participant nucleons, similar to {\em nuclear transparency} as observed for fixed-target \pa\ experiments in the seventies.
(d) The same analysis applied to \ppb\ spectra reveals  that a similar scenario applies, with minor quantitative differences due to the different collision geometries.
(e) A collision scenario involving formation of a high-density flowing QCD medium appears inconsistent with those findings.
It seems ironic that the apparently strongest diagnostic for jet suppression or quenching by a dense flowing QCD medium actually provides the strongest evidence for nuclear transparency (via exclusivity and time dilation).


\end{document}